\documentclass[%
 reprint,
nofootinbib,
 amsmath,amssymb,
 aps,
]{revtex4-2}
\usepackage{graphicx}% Include figure files
\usepackage{dcolumn}% Align table columns on decimal point
\usepackage{bm}% bold math
\usepackage{lipsum} 
\usepackage{tikz}
\usetikzlibrary{shapes.geometric, arrows, backgrounds, calc, fit}
\usepackage{layouts}
\usepackage{float}
\usepackage{amsmath}
\usepackage{dsfont}
\usepackage{physics}
\usepackage{graphicx}
\usepackage{enumitem}
\usepackage{bbold}
\usepackage{xcolor}
\usepackage[colorlinks=true, allcolors=blue]{hyperref}
\usepackage{siunitx}
\usepackage[bottom]{footmisc}
\usepackage{todonotes}
\usepackage{booktabs}
\usepackage{multirow}
\usepackage{nicematrix}

\newif\ifdraft
\drafttrue
\ifdraft
\newcommand{\note}[1]{ {\textcolor{orange} { **Note: #1 }}}

\newcommand{\alnote}[1]{ {\textcolor{blue} { ***Andre: #1 }}}
\newcommand{\lhnote}[1]{ {\textcolor{orange} { ***Leo: #1 }}}

\else
\newcommand{\note}[1]{}
\newcommand{\alnote}[1]{}
\newcommand{\lhnote}[1]{}
\fi
\newcommand{\ignore}[1]{}

\begin{document}

% \title{Quantum-Inspired Tensor Network Algorithm for Computational Fluid Dynamics with GPU-Accelerated Computations}
\title{Quantum-Inspired Fluid Simulation of 2D Turbulence with GPU Acceleration}
% \title{Quantum-Inspired Simulations of Turbulent Flows with GPU Acceleration}
% \title{High-Order Time Stepping and GPU Acceleration in Quantum-Inspired Turbulence Simulation}

\author{Leonhard Hölscher$^{1,2,3}$}
\email{leonhard.hoelscher@bmw.de}
\author{Pooja Rao$^{4}$}
\author{Lukas Müller$^{1}$}
\author{Johannes Klepsch$^{1}$}
\author{Andre Luckow$^{1,5}$}
%\author{Jin-Sung Kim$^{4}$}
\author{Tobias Stollenwerk$^{3}$}
\author{Frank K. Wilhelm$^{2,3}$}
\affiliation{$^{1}$BMW Group, Munich, Germany}
\affiliation{$^{2}$Theoretical Physics, Saarland University, 66123 Saarbrücken, Germany}
\affiliation{$^{3}$Institute for Quantum Computing Analytics (PGI 12),
Forschungszentrum Jülich, 52425 Jülich, Germany}
\affiliation{$^{4}$Quantum Algorithm Engineering, NVIDIA, Santa Clara, CA, USA}
\affiliation{$^{5}$Ludwigs Maximilian University Munich, Munich, Germany}

\date{\today}

\begin{abstract}
Tensor network algorithms can efficiently simulate complex quantum many-body systems by utilizing knowledge of their structure and entanglement. These methodologies have been adapted recently for solving the Navier-Stokes equations, which describe a spectrum of fluid phenomena, from the aerodynamics of vehicles to weather patterns. Within this quantum-inspired paradigm, velocity is encoded as matrix product states (MPS), effectively harnessing the analogy between interscale correlations of fluid dynamics and entanglement in quantum many-body physics. This particular tensor structure is also called quantics tensor train (QTT). By utilizing NVIDIA's cuQuantum library to perform parallel tensor computations on GPUs, our adaptation speeds up simulations by up to 12.1 times. This allows us to study the algorithm in terms of its applicability, scalability, and performance. By simulating two qualitatively different but commonly encountered 2D flow problems at high Reynolds numbers up to $\SI{1e7}{}$ using a fourth-order time stepping scheme, we find that the algorithm has a potential advantage over direct numerical simulations in the turbulent regime as the requirements for grid resolution increase drastically. In addition, we derive the scaling $\chi=\mathcal{O}(\text{poly}(1/\epsilon))$ for the maximum bond dimension $\chi$ of MPS representing turbulent flow fields, with an error $\epsilon$, based on the spectral distribution of turbulent kinetic energy. Our findings motivate further exploration of related quantum algorithms and other tensor network methods.

\end{abstract}

\maketitle

%\tableofcontents
% \onecolumngrid
\section{Introduction}
Tensor network (TN) algorithms \cite{verstraete_matrix_2008, bridgeman_hand-waving_2017} play a crucial role in simulating complex quantum many-body systems by utilizing knowledge of their structure and entanglement to provide accurate approximations. Among the most notable of these algorithms is the density-matrix renormalization group (DMRG) method \cite{white_density_1992, white_density-matrix_1993, schollwock_density-matrix_2005, schollwock_density-matrix_2011}, a powerful technique tailored for simulating one-dimensional quantum spin models. Although these methods were initially designed to tackle specific linear algebra problems within quantum physics, their utility spans wider, inspiring the development of algorithms in other areas, now referred to as quantum-inspired algorithms \cite{garcia-ripoll_quantum-inspired_2021}.

One of these areas includes computational fluid dynamics (CFD) \cite{ferziger_computational_2020} employing numerical methods to solve the Navier-Stokes equations, which describe a wide range of fluid phenomena, from the aerodynamics of vehicles to weather patterns. In general, fluid simulations are computationally expensive as the Navier-Stokes equations are a complex set of coupled and nonlinear partial differential equations (PDEs) \cite{fefferman2000existence}. The computational demand escalates with increasing turbulent behavior \cite{pope_turbulent_2000, nichols2010turbulence}. To tackle this, several strategies have been developed, including Large Eddy Simulations (LES) \cite{galperin_large_2010} or implicit LES (ILES) \cite{ferziger_computational_2020}, each offering a distinct approach to capture turbulence effectively.

Recent research has explored the use of quantum-inspired algorithms for solving the incompressible Navier-Stokes equations \cite{gourianov_quantum-inspired_2022, gourianov_exploiting_2022}, with Gourianov et al. pioneering this approach by encoding the velocity field as a matrix product state (MPS), a prevalent TN also known as tensor train (TT) \cite{oseledets_tensor-train_2011}. This method draws an analogy between the local correlations of quantum states, known as entanglement, and the correlations of length scales in turbulent flows. The concept mirrors the area law observed in quantum states \cite{eisert_colloquium_2010}, suggesting that these correlations are limited as interactions predominantly occur between flow structures of similar length scales. This idea aligns with the Kolmogorov-Richardson energy cascade theory for 3D turbulence \cite{richardson_weather_2007, kolmogorov_local_1941}, which outlines the gradual transfer of kinetic energy down to smaller scales until dissipation into heat at the Kolmogorov scale. A similar multiscale view is valid for 2D turbulence, although it contains a different energy cascade mechanism \cite{kraichnan_inertial_1967, batchelor_computation_1969, leith_atmospheric_1971}.

The quantum-inspired CFD algorithm indicated a potential advantage over direct numerical simulation (DNS) of 2D decaying jet flow \cite{gourianov_quantum-inspired_2022} by truncating length scale correlations. This suggests that quantum-inspired approaches could be a candidate to tackle highly turbulent simulations. A similar advantage has not been observed yet for 3D simulations.

Related work has shown that quantum-inspired Navier-Stokes solvers can be extended to handle complex boundaries \cite{kiffner_tensor_2023, kornev_numerical_2023} and to incorporate immersed objects via masking techniques \cite{peddinti_quantum-inspired_2024}. These advancements are crucial for their adoption in real-world scientific and engineering problems, which often involve intricate shapes and boundary conditions. 

In this work, we examine various aspects of the method proposed by Gourianov et al. \cite{gourianov_quantum-inspired_2022} and augment it with a fourth-order time-stepping scheme to reduce numerical errors. We demonstrate a significant speedup in simulations through parallelized tensor network operations on GPUs. This advancement enables a more thorough empirical analysis than previously presented. For instance, we explore the relationship between the maximum bond dimension $\chi$ — essential for managing the truncation of correlations across different length scales — and its behavior with increasing Reynolds numbers up to $\mathrm{Re}=\SI{1e7}{}$, which is two orders of magnitude larger than what was previously studied in this context. This investigation is critical to assessing the method's scalability and effectiveness in addressing fluid dynamics problems with varying turbulence levels, as quantified by the Reynolds number. In addition to studying the anisotropic jet flow from \cite{gourianov_quantum-inspired_2022}, we also showcase the application of this algorithm to a 2D isotropic decaying turbulence flow. Moreover, we present the resource requirements of the method from a practical standpoint and identify scenarios where an advantage from memory compression could be realized. One of our main contributions lies in giving a theoretical explanation why the MPS format gives an efficient approximation for turbulent flow fields. Finally, we compare the algorithm's results to DNS by calculating pointwise fidelities and turbulence kinetic energy spectra in wavespace, commonly used in turbulence analysis, to evaluate the multiscale flow behavior. Our findings are significant for the further development of this emerging area of research and provide valuable insights for TN algorithms and related quantum algorithms \cite{lubasch_variational_2020, jaksch_variational_2022}. Our paper also clarifies some of the implementation details of the algorithm that were not covered in the reference work and also provides details on a more memory-efficient and GPU parallelized version. For easy reproducibility, the DNS and the MPS codes used in this paper alongside a tutorial have been made open source \cite{Hoelscher_TN_CFD}.

The remainder of the paper is organized as follows. Sec.\,\ref{sec:methods} is devoted to the introduction of incompressible fluid dynamics and the employed TN methods to solve them. In Sec.\,\ref{sec:results} we analyze the TN method and its implications for the two flow problems under study. We conclude our findings and give an outlook by suggesting potential improvements for future quantum-inspired CFD solvers in Sec.\,\ref{sec:discussion}.

\section{Methods}\label{sec:methods}
\subsection{Incompressible Fluid Dynamics}\label{sec:ifd}
The Navier-Stokes equations govern the fundamental principles of fluid dynamics. Specifically, these equations describe the conservation of mass and momentum for incompressible fluids with constant density through a system of coupled partial differential equations \cite{ferziger_computational_2020}. The continuity equation is described as
\begin{equation}
    \nabla\cdot\bm{u} = 0
    \label{eq:continuity}
\end{equation}
and the momentum equation is
\begin{equation}
    \frac{\partial}{\partial t}\bm{u} = \underbrace{-\left(\bm{u}\cdot\nabla\right)\bm{u}}_\text{convection} + \underbrace{\frac{1}{\text{Re}}\nabla^2 \bm{u}}_\text{diffusion} - \nabla p \text{.}
    \label{eq:momentum}
\end{equation}
Here, $\bm{u}$ is the velocity vector, $p$ is the pressure, and $\text{Re}$ is the Reynolds number. In this formulation, all variables are non-dimensionalized, and the Reynolds number indicates how turbulent the flow is. Furthermore, the momentum equation can be split into convection, diffusion, and pressure terms, each describing individual physical phenomena.
\subsection{Algorithmic Building Blocks}\label{sec:encoding}\label{sec:diff_mpos_theory}
Here, we introduce several algorithmic building blocks necessary for quantum-inspired fluid simulations, including \textit{MPS encoding}, \textit{differential operators}, and \textit{nonlinear operations} as required by the convection term in Eq.\eqref{eq:momentum}.
\\

\begin{figure}[H]
    \centering
    \includegraphics[]{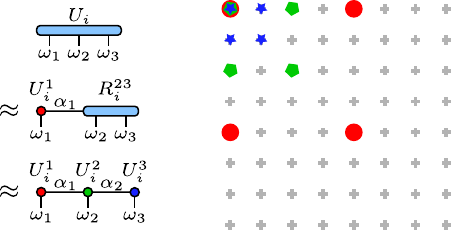}
    \caption{Illustration of the \textit{quantics} representation of each velocity component $U_i^{\omega_1\omega_2\omega_3}$ and its MPS decomposition using three bits per spatial dimension. Each tensor corresponds to the part of the velocity residing on a $2\times2$ grid of individual length scale.}
    \label{fig:length_scale_encoding}
\end{figure}

In this work, we focus on 2D flows, though, the methods outlined here can be generalized to 3D. The main quantity of interest is the velocity $\bm{u}=(u_1, u_2)$, discretized on a uniform $N \times N$ grid where $N=2^n$ and $n$ is the number of bits per spatial component $u_i(x_1, x_2)$. Here, $x_i$ is an index going from 0 to $N-1$. Each velocity component can thus be written as a rank-2 tensor $U_i^{x_1x_2}$. Due to the multiscale nature of turbulent flows \cite{pope_turbulent_2000}, we chose the so-called \textit{quantics} representation \cite{khoromskij_odlog_2011, lindsey_multiscale_2024, ritter_quantics_2024} for the velocity tensors as it naturally encodes scale separation. This entails splitting the indices corresponding to spatial coordinates $x_1$ and $x_2$ in their binary components $(x_i^1x_i^2\dots x_i^n)_2$ and rearranging them to form new multi-indices $\omega_k = (x_1^kx_2^k)_2$.
\begin{align}
\begin{split}
    U_i^{x_1x_2} &= \:
    \begin{gathered}
        \includegraphics[scale=1.3]{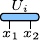}
    \end{gathered}
    \:=\:
    \begin{gathered}
        \includegraphics[scale=1.3]{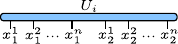}
    \end{gathered}
    \\
    &=\:
    \begin{gathered}
        \includegraphics[scale=1.3]{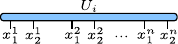}
    \end{gathered}
    \\
    &=\:
    \begin{gathered}
        \includegraphics[scale=1.3]{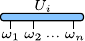}
    \end{gathered}
    \:= U_i^{\omega_1\omega_2\dots\omega_n}
\end{split}
\end{align}
This yields a velocity tensor $U_i^{\omega_1\omega_2\dots\omega_n}$,  where each $\omega_k$ corresponds to a particular length scale. The intuition is that we can efficiently approximate the velocity as an MPS with maximum bond dimension $\chi$
\begin{equation}
    U_i^{\omega_1\omega_2\ldots\omega_{n}} \approx \sum_{\{\alpha_\ell\}=1}^{\leq\chi} U^{\omega_1}_{i, \alpha_1} U^{\omega_2}_{i, \alpha_1\alpha_2} \ldots U^{\omega_{n}}_{i, \alpha_{n-1}}
\end{equation}
because the correlations between length scales are limited.
This allows us to express $U_i^{\omega_1\omega_2\ldots\omega_{n}}$ with $\mathcal{O}(n\chi^2)$ instead of $\mathcal{O}(4^n)$ parameters.
Fig.\,\ref{fig:length_scale_encoding} illustrates the MPS decomposition of the velocity field using three bits per spatial dimension and highlights the corresponding $2\times2$ subgrids of different length scales defined by $\omega_k$. 
Each tensor of the MPS describes a part of the velocity living on the respective subgrid. In analogy to quantum mechanics, we can also interpret the tensor $U_i^{\omega_1\omega_2\ldots\omega_{n}}$ as a quantum state
\begin{equation}
    \ket{u_i} = \sum_{\{\omega_k\}=1}^4 U_i^{\omega_1\omega_2\ldots\omega_{n}} \ket{\omega_1\omega_2\ldots\omega_{n}}\text{.}
\end{equation}
The implications of this encoding and the required $\chi$ are analyzed empirically in Sec.\,\ref{sec:approx} and theoretically in Sec.\,\ref{sec:derivation}.
\\

In order to solve differential equations, we need differential operators that can be applied to MPS representing the velocity. Matrix Product Operators (MPO) are the natural choice. They have a similar structure as an MPS and act on them by contracting the respective indices. One can construct MPOs to perform simple arithmetic operations such that a derivative can be approximated using finite differences. For example, the simple central finite difference operator $\hat{D}_1$ with respect to $x_1$ yields
\begin{equation}
    \frac{\partial U_i^{x_1x_2}}{\partial x_1} \approx \frac{U_i^{(x_1+1)x_2}}{2\Delta x} - \frac{U_i^{(x_1-1)x_2}}{2\Delta x} \text{.}
    \label{eq:fdm}
\end{equation}
Here, $\Delta x = 1/(N-1)$ is the spacing between two adjacent grid points.
Consequently, $\hat{D}_1$ computes this sum for each grid point and can be constructed from a repetitive tensor structure:
\begin{equation}
    \hat{D}_1 = 
    \begin{gathered}
        \includegraphics[scale=1.3]{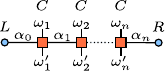}
    \end{gathered}
    \mathrm{.}
    \label{eq:mpo_d1}
\end{equation}
The boundary tensors $L_{\alpha_0}$ and $R_{\alpha_n}$ define the periodic boundary conditions and the coefficients in Eq.\eqref{eq:fdm}, respectively. The central rank-4 tensors $C_{\alpha_{k-1}\alpha_k}^{\omega_k\omega_k'}$ are identical and contain only $1$'s for mappings to itself and neighboring grid points via binary addition and subtraction while every other value is set to $0$. More technical details and exact tensor values of $\hat{D}_1$ are given in Appendix\,\ref{sec:diff_mpos}. Using the same logic, one can build finite difference operators of any order \cite{garcia-ripoll_quantum-inspired_2021, kazeev_low-rank_2012}. 
\\

\begin{figure}[]
    \centering
    \includegraphics{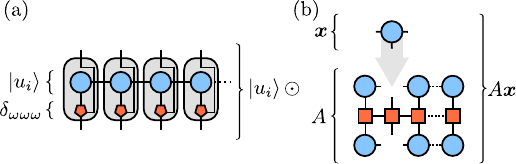}
    \caption{Algorithmic building blocks. (a) Construction of the MPO $\ket{u_i}\odot$ for the nonlinear operation $\ket{u_i}\odot\ket{u_j}$ using rank-3 Kronecker delta tensors. (b) Matrix vector multiplication $A\bm{x}$ during the DMRG-like optimization routine.}
    \label{fig:nonlinear_Ax}
\end{figure}

A fundamental difficulty of solving the Navier-Stokes equations is the nonlinear convective term in Eq.\,\eqref{eq:momentum}. While solving nonlinear problems on quantum computers is still an ongoing research field \cite{liu_efficient_2021, joseph_koopman-von_2020, lloyd_quantum_2020, kyriienko_solving_2021, lubasch_variational_2020}, nonlinear TN problems can be solved with reasonable accuracy with classical computers. By using repeated copy operations and neglecting the need for a normalized state, one can naively construct an element-wise product of two states 
\begin{equation}
    \ket{u_i\cdot u_j} = \ket{u_i}\odot\ket{u_j}\text{.}
\end{equation}
Here, $\ket{u_i}\odot$ can be interpreted as an MPO acting on $\ket{u_j}$. This operator corresponds to $\ket{u_i}$, where each tensor is contracted with rank-3 Kronecker delta tensors $\delta_{\omega\omega\omega}$. Fig.\,\ref{fig:nonlinear_Ax}(a) shows the diagrammatic construction of this operator. The resulting MPS $\ket{u_i\cdot u_j}$ has a maximum bond dimension of $\chi^2$ and needs to be compressed down to $\chi$ again.

\subsection{Quantum-Inspired CFD Algorithm}\label{sec:QIS}
\begin{figure*}
    \centering
    \includegraphics[width=\textwidth]{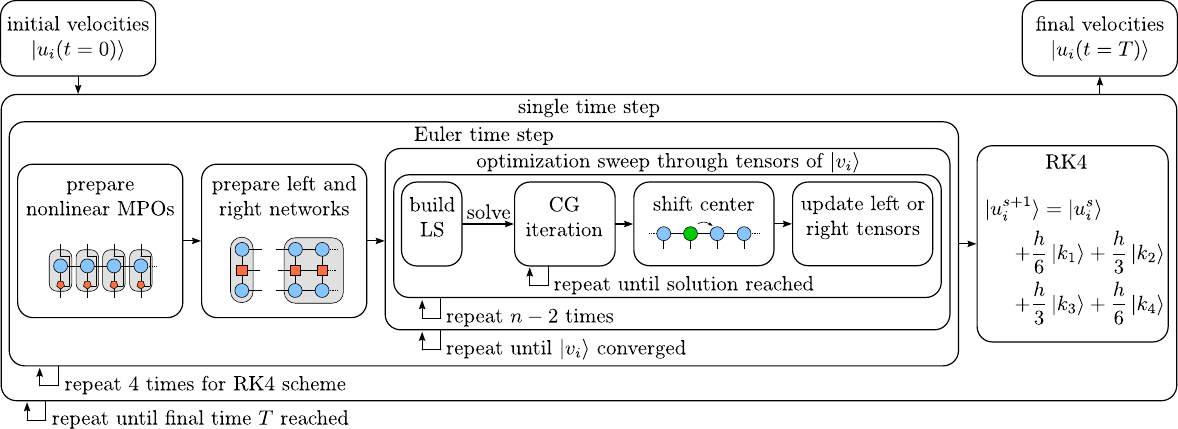}
    \caption{Schematic overview of the quantum-inspired CFD algorithm.}
    \label{fig:flow_chart}
\end{figure*}
We employ a modified version of the algorithm proposed by \cite{gourianov_quantum-inspired_2022}, which aims to solve the incompressible Navier-Stokes equations in variational form. Similar to how the DMRG method \cite{verstraete_matrix_2008, bridgeman_hand-waving_2017} finds the ground state of a system described by a Hamiltonian $\hat{H}$ by minimizing the energy $\bra{\psi}\hat{H}\ket{\psi}$ of a trial state $\ket{\psi}$, we find the velocity states $\ket{u_i^{s+1}}$ for the subsequent time step $s+1$ by minimizing a cost function $\Theta\left(\ket{v_1}, \ket{v_2}, \Delta t\right)$ based on the Navier-Stokes equations. Here, $\ket{v_i}$ are the trial states and $\Delta t$ corresponds to the spacing in time $t_{s+1}-t_s$. By solving
\begin{equation}
    \left(\ket{u_1^{s+1}}, \ket{u_2^{s+1}}\right) = \underset{\ket{v_1}, \ket{v_2}}{\arg\min}\ \Theta\left(\ket{v_1}, \ket{v_2}, \Delta t\right)\text{,}
    \label{eq:minimization}
\end{equation}
the explicit Euler method is employed to step forward in time:
\begin{align}
\begin{split}
    \ket{u_i^{s+1}} &= \ket{u_i^{s}} + \Delta t\cdot \ket{f\left(t_s, \ket{u_1^{s}}, \ket{u_2^{s}}\right)}\\ \text{where}\quad\frac{d\ket{u_i}}{dt} &\approx \ket{f\left(t, \ket{u_1}, \ket{u_2}\right)}\mathrm{.}
\end{split}
\end{align}
Without considering errors from the finite difference approximation of the spatial derivatives, the solution of the Euler method has an error of $\mathcal{O}(\Delta t^2)$. To be able to simulate turbulent flows with Reynolds numbers up to $\mathrm{Re}=\SI{1e7}{}$, we reduce this error to $\mathcal{O}(\Delta t^5)$ by employing the fourth order Runge-Kutta method (RK4). Thus, we perform four minimizations as described in Eq.\,\eqref{eq:minimization} to calculate $\ket{u_i^{s+1}}$. This becomes clear by interpreting the RK4 step \cite{ferziger_computational_2020}
\begin{equation}
    \ket{u_i^{s+1}} = \ket{u_i^{s}} + \frac{\Delta t}{6}\left(\ket{k_1}+2\ket{k_2}+2\ket{k_3}+\ket{k_4}\right)
\end{equation}
as four distinct Euler steps. The minimization of $\Theta$ is done by iteratively updating the tensors of the trial MPS $\ket{v_i}$. Each tensor is updated by solving a particular linear system (LS) of equations $A\bm{x}=\bm{b}$, where $A$ and $\bm{x}$ correspond to tensors as depicted in Fig.\,\ref{fig:nonlinear_Ax}(b). The derivation of $\Theta$, its exact minimization procedure, and the computation of the RK4 gradients $\ket{k_{j}}_{j\in\{1, 2, 3, 4\}}$ is explained in Appendix \ref{sec:cost}.

The entire workflow of the DMRG-like algorithm is sketched in Fig.\,\ref{fig:flow_chart}. The initial velocities as MPSs at time $t=0$ are given as input to the algorithm. Then, we update the velocities for a single time step and repeat this computation until we have reached the desired final time $t=T$ and return the velocities as output. A single time step of RK4 consists of a sum of four Euler time steps. This addition step is dominated by the compression of the MPSs back down to a maximum bond dimension of $\chi$, which is done using repeated singular value decompositions (SVDs) with typical complexity of $\mathcal{O}(\chi^3)$ \cite{schollwock_density-matrix_2011}. The Euler time step begins with preparing the nonlinear MPOs with complexity $\mathcal{O}(n\chi)$. This is followed by contracting left and right TNs for the iterative DMRG-like optimization process. This step is computationally expensive as it contains the contraction of the nonlinear MPOs with the velocity states, both of which have a maximum bond dimension $\chi$. Thus, this step has complexity $\mathcal{O}(n\chi^4)$. Then, we sweep through the tensors of the trial velocity states, which are optimized and updated one by one. Therefore, we begin the sweep in the canonical center of the trial MPSs and build the LS according to the cost function $\Theta$. This is solved using the conjugate gradient (CG) algorithm, where a single iteration has complexity $\mathcal{O}(\chi^3)$. In our implementation, the number of CG iterations is limited to a maximum of 100, but it can be lower if the residual is smaller than the tolerance of $\SI{1e-5}{}$. Once we have found the new tensor of the MPSs, we shift the canonical center to the next adjacent tensor using SVDs ($\mathcal{O}(\chi^3)$). Finally, we must update the left or right TNs that were contracted in the beginning. This update step scales as $\mathcal{O}(\chi^4)$ due to the nonlinear MPOs. The optimization sweep ends once the canonical center is back at its initial position and the trial state has converged. Convergence is achieved when the relative change of the sum of the inner products of the velocity components is less than a tolerance of $\SI{1e-5}{}$. Hence, the overall complexity of the algorithm is $\mathcal{O}(n\chi^4)$ \cite{gourianov_quantum-inspired_2022, gourianov_exploiting_2022}.

\subsection{Efficient Implementation}
An efficient implementation of TN algorithms is pivotal for leveraging their complexity advantages. Hence, special attention must be paid to intermediate contractions, as exemplified when computing $A\bm{x}$ and depicted in Fig.\,\ref{fig:nonlinear_Ax}(b). Directly contracting $A$ would yield a large tensor, with all six indices contributing to its size. Instead, maintaining $A$ in its decomposed form and solely executing contractions for computing $A\bm{x}$ is more efficient.

Moreover, determining the optimal contraction path presents an optimization problem in itself. However, given that the dimensions of all tensors are known by setting the maximum bond dimension $\chi$, it is computationally more efficient to explicitly define the optimal path for all recurring contractions.

As contractions constitute a significant aspect of TN algorithms, harnessing GPUs for acceleration is a natural choice. However, the limited memory of GPUs underscores the importance of efficiently managing workspace memory. This can be accomplished adeptly through the \verb|cuQuantum| library \cite{bayraktar_cuquantum_2023}.
% The implementation of all algorithms used for this work is accessible in \cite{Hoelscher_TN_CFD}.

\section{Results}\label{sec:results}
\begin{figure}[H]
    \includegraphics{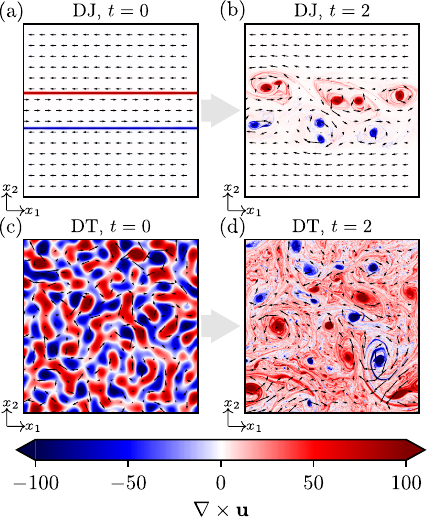}
    \caption{Analyzed flows. Arrows represent the velocity field $\bm{u}$, and the colormap illustrates the vorticity $\nabla\times\bm{u}$. Panel (a) depicts the initial conditions of the decaying jet (DJ) flow, while panel (b) shows its state at $t=2$. Panels (c) and (d) similarly illustrate the decaying turbulence (DT) flow at $t=0$ and $t=2$, respectively. Both simulations used a Reynolds number of $\mathrm{Re}=\SI{1e7}{}$}
    \label{fig:problems}
\end{figure}

This section provides a thorough numerical analysis of the quantum-inspired CFD algorithm, along with theoretical insights into MPS encoding.
As illustrated in Fig.\,\ref{fig:problems}, we examine two distinct flows, namely the decaying jet (DJ) \cite{gourianov_quantum-inspired_2022} and the decaying turbulence (DT) \cite{san_high-order_2012} problems. 
The DJ flow is initialized as a horizontal jet, and exhibits Kelvin-Helmholtz instabilities at the shear layers over time. On the other hand, the DT flow is randomly initialized, creating an isotropic chaotic behavior (cf. Appendix\,\ref{sec:initial}). Both flows have periodic boundary conditions and evolve without any external forces. 
\subsection{Algorithm Verification}\label{sec:verification}
\begin{figure}[H]
    \centering
    \includegraphics{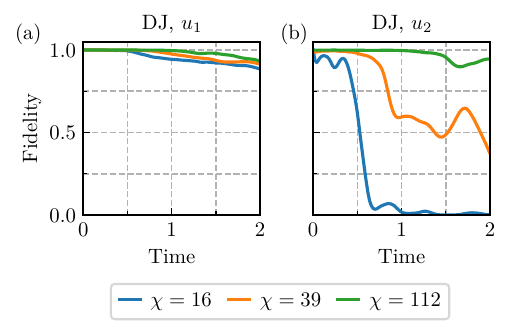}
    \caption{Fidelities according to Eq.\eqref{eq:fidelity} between DNS and QIS for different $\chi$. Panels (a) and (b) show the fidelities for the $u_1$ and $u_2$ components of DJ simulations with $\mathrm{Re}=\SI{2e5}{}$.}
    \label{fig:fidelities_low_Re}
\end{figure}

To verify the algorithm, we compare the results from quantum-inspired simulation (QIS) with direct numerical simulation (DNS). The DNS scheme is based on finite differences and the Fast Fourier transforms as explained in Appendix \ref{sec:dns}. We use the quantum fidelity 
\begin{equation}
    F = \frac{|\langle u_i^\mathrm{DNS}|u_i^\mathrm{QIS}\rangle|^2}{\lVert|u_i^\mathrm{DNS}\rangle\rVert_2^2\lVert|u_i^\mathrm{QIS}\rangle\rVert_2^2}
    \label{eq:fidelity}
\end{equation}
as a metric to compare the results over time between the QIS and DNS methods. Fig.\,\ref{fig:fidelities_low_Re} shows the fidelity over time for the velocity components $u_1$ and $u_2$ of DJ simulations with $n=10$, $\Delta t=0.1/2^9$, and different $\chi$. 
\noindent The fidelity of the $u_1$ component remains close to $1$ for all tested $\chi$ values, indicating high accuracy. However, the fidelity of the $u_2$ component decreases significantly over time for $\chi=16$ or $\chi=39$. For $\chi=112$, the captured number of correlations is sufficient to achieve high accuracy for $u_2$ as well. The observed anisoropic behavior is attributed to the initial conditions of the DJ flow, where the velocity initially points either in positive or negative $x_1$ direction. Over time, Kevin-Helmholtz instabilities appear at the shear layers, causing a comparatively significant increase in $u_2$, while $u_1$ remains relatively unaffected. Consequently, $u_2$ requires a larger $\chi$ to reach the same level of accuracy as $u_1$, as demonstrated in Sec.\,\ref{sec:approx}. 

\begin{figure}[h]
    \centering
    \includegraphics{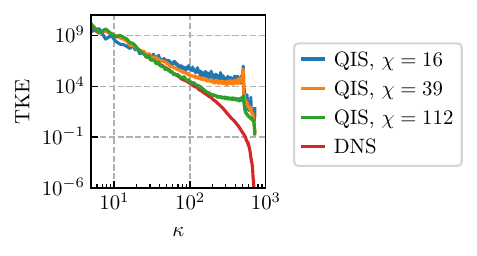}
    \caption{TKE spectrum for the DJ flow with $\mathrm{Re} = \SI{2e5}{}$ at $t=2$. }
    \label{fig:TKE_low_Re}
\end{figure}

Since real-world turbulence is a statistical phenomenon \cite{pope_turbulent_2000} and the fidelity is a measure for point-wise agreement, we also examine and qualitatively compare the turbulent kinetic energy (TKE) spectra in wave space. The TKE spectrum is calculated as
\begin{equation}
    E(\kappa_1, \kappa_2) = \frac{1}{2} \left(\hat{u}_1'(\kappa_1, \kappa_2)^2 + \hat{u}_2'(\kappa_1, \kappa_2)^2\right)\mathrm{,}
    \label{eq:TKE}
\end{equation}
where $\hat{u}_i'(\kappa_1, \kappa_2)$ is the Fourier transform of the fluctuating part of the instantaneous velocity $u_i(x_1, x_2)$, which can be extracted from the Reynolds decomposition of the instantaneous velocity $u_i$:
\begin{equation}
    u_i' = u_i - u_i^\text{mean}\mathrm{.}
\end{equation}
Here, $u_i^\text{mean}$ is the velocity component averaged over time. The TKE spectrum against wave number $\kappa = \sqrt{\kappa_1^2+\kappa_2^2}$ illustrates the distribution of kinetic energy across different length scales represented by $\kappa$ \cite{pope_turbulent_2000, kraichnan_inertial_1967, batchelor_computation_1969, leith_atmospheric_1971} and can be calculated as
\begin{equation}
    E(\kappa) = \iint_{\kappa^2 = \kappa_1^2+\kappa_2^2} E(\kappa_1, \kappa_2) d\kappa_1d\kappa_2\mathrm{.}
\end{equation}
Fig.\,\ref{fig:TKE_low_Re} shows the TKE spectrum for QIS and DNS of the DJ flow at time $t=2$.
At smaller wave numbers, QIS results align with DNS results, but significant discrepancies emerge at higher $\kappa$. QIS seemingly leads to an accumulation of kinetic energy at small length scales compared to DNS. This discrepancy arises from MPS compression, as the error increases with smaller $\chi$. MPS is effective in approximating functions with rapidly decaying Fourier coefficients \cite{lindsey_multiscale_2024, jobst_efficient_2023, dolgov_fast_2012}. Since the TKE is directly proportional to the Fourier coefficients of the velocity, the error is predominantly observed at large wave numbers.

\subsection{Runtime Analysis}\label{sec:runtime}
\begin{figure}[H]
    \centering
    \includegraphics{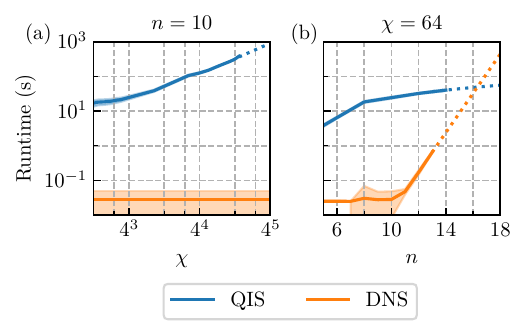}
    \caption{Measured runtime for simulating a single time step using quantum-inspired simulation (QIS) and direct numerical simulation (DNS) of the DT flow on GPUs. Panel (a) shows the dependence on $\chi$, while panel (b) shows the dependence on $n$. Shaded areas indicate the standard deviation from 100 repetitions, and dotted lines show the expected extrapolated behavior.}
    \label{fig:runtime}
\end{figure}

To assess the practical runtime of the algorithm, we conducted runtime experiments for QIS and DNS of the DT flow on our system (CPU: Intel Xeon Platinum 8480CL with 2 TB RAM; GPU: NVIDIA H100 with 80 GB). For executing the QIS method on CPU, we utilized the Python library \verb|quimb| \cite{gray2018quimb}, and for the GPU version, we leveraged the Python API of \verb|cuQuantum| \cite{bayraktar_cuquantum_2023}. Fig.\,\ref{fig:runtime} shows the measured runtime for simulating a single time step of the DT flow using the QIS and the DNS algorithms on GPUs. 
\noindent The runtime has been averaged over 100 time steps, starting at $t=1$, to ensure that the flow significantly diverged from its initial conditions and the simulation has fully developed. The DNS runtime is unaffected by $\chi$ as it always computes the most accurate solution and, as a result, remains constant. The QIS, on the other hand, scales polynomially with $\chi$. As explained in Sec.\,\ref{sec:QIS}, the QIS has a theoretical time complexity of $\mathcal{O}(n\chi^4)$, however, we observe an empirical scaling $\propto\chi^{1.5}$. This polynomial runtime reduction between theory and practice is likely due to the efficient implementation of tensor contractions which are well-parallelizable. For a grid size corresponding to $n=10$, it is apparent that QIS shows no speedup compared to DNS. However, this changes when the runtime is compared with increasing 
$n$. For a fixed $\chi=64$, the QIS shows expected linear scaling with $n$, while the DNS illustrates an exponential runtime scaling in accordance with its computational complexity of $\mathcal{O}(4^nn)$. Due to large memory requirements, the DNS runs have been restricted to $n<14$. In this regime, the QIS shows no runtime advantage over DNS. However, as both the empirical trend and the theoretical complexity argument suggest, the advantage over DNS is likely to appear for larger $n$. By extrapolating our runtime data, a computational advantage is expected for $n>16$. 

\begin{figure}[]
    \centering
    \includegraphics{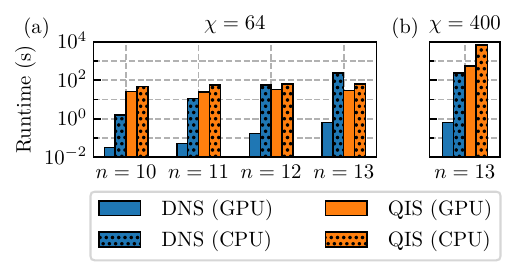}
    \caption{Runtime comparison of CPU and GPU simulations of a single time step of the DT flow for $\chi=64$ and varying $n$ in panel (a) and $\chi=400$ and $n=13$ in panel (b).}
    \label{fig:cpu_gpu_runtime}
\end{figure}

As part of the comparison between QIS and DNS, we also evaluate the runtimes of our GPU and CPU implementations, as shown in Fig.\,\ref{fig:cpu_gpu_runtime}. Our GPU implementation consistently outperforms its CPU counterpart, highlighting the effectiveness of GPUs for such methods. Specifically, we observe a 383-fold speedup for DNS and a 2.2-fold speedup for QIS for $n=13$ and $\chi=64$. While the computation time for QIS has been halved using GPUs for this configuration, the speedup may seem small compared to the DNS speedup. However, the QIS speedup increases significantly with increasing $\chi$. For instance for $n=13$ and $\chi=400$, we already see a 12.1-fold speedup for QIS. If we only compare CPU runtimes, our data reveals that runtime advantage of QIS with $\chi=64$ over DNS becomes apparent for $n>12$. For larger $\chi$, this cross-over point shifts upwards to larger $n$.

\subsection{Memory Consumption}\label{sec:memory}
\begin{figure}[]
    \centering
    \includegraphics{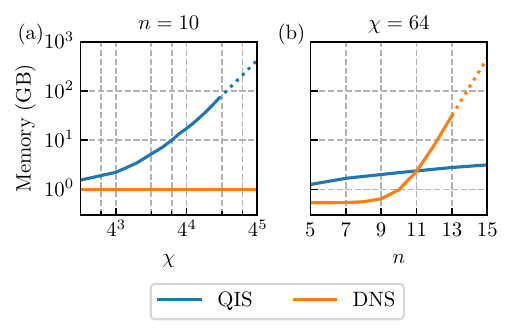}
    \caption{Occupied memory of QIS and DNS algorithms. Panel (a) shows the memory consumption for $n=10$ and varying $\chi$, and panel (b) shows it for $\chi=64$ and varying $n$. As our GPU-memory is limited to $\SI{80}{\giga\byte}$, the dotted lines beyond that are extrapolated.}
    \label{fig:memory}
\end{figure}

The number of parameters of an MPS scales as $\mathcal{O}(n\chi^2)$, which is an exponential improvement with respect to $n$ compared to the full tensor scaling of $\mathcal{O}(4^n)$. We measure the QIS algorithm's memory usage to determine if the reduced number of parameters is reflected in its memory footprint. Fig.\,\ref{fig:memory}(a) demonstrates the relationship between the memory requirement and $\chi$ for QIS and DNS. Similar to the previously presented runtime analysis, the DNS data appears as a constant line as it is independent of $\chi$. In contrast, the QIS scales quadratically with $\chi$. Fig.\,\ref{fig:memory}(b) shows the memory consumption as a function of $n$. Here, the QIS demonstrates perfect linear scaling, whereas the DNS showcases exponential scaling. Thus, the QIS's memory requirements reflect the compression advantage associated with MPS. However, our data for $\chi=64$ shows that this memory advantage is only realized for $n>11$. Again, larger $\chi$ would shift the cross-over point further to larger $n$. 

\subsection{Grid Convergence Study}
\begin{figure}[h]
    \centering
    \includegraphics{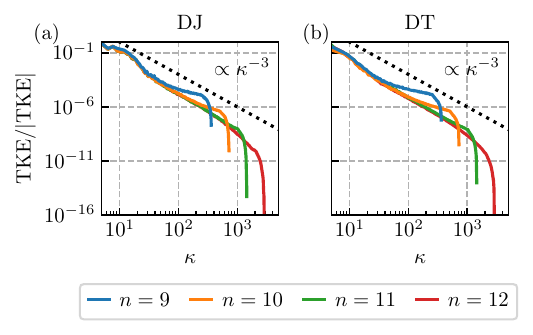}
    \caption{Grid convergence study using normalized turbulent kinetic energy (TKE) spectra from DNSs of (a) DJ and (b) DT with $\mathrm{Re}=\SI{1e7}{}$ at $t=2$. The dashed line illustrates the theoretical scaling of $\propto\kappa^{-3}$ in the inertial range for infinite $\mathrm{Re}$.}
    \label{fig:grid_convergence}
\end{figure}

For most flows of interest, direct numerical simulations of turbulent flows require a prohibitively high grid resolution to resolve all relevant length scales, from the largest energy-producing scales $\ell_0$ to the smallest dissipation scales $\eta$. For these fully resolved DNS simulations, the number of grid points per spatial dimension should satisfy $N=2^n\geq\ell_0/\eta$, which is directly proportional to the Reynolds number $\ell_0/\eta\propto\mathrm{Re}$ for 2D\footnote{For 3D turbulence, we have $\ell_0/\eta\propto\mathrm{Re}^{3/4}$ \cite{pope_turbulent_2000}.} turbulence \cite{lesieur_turbulence_2008}. We conduct a grid convergence study to verify that our chosen grid resolution is sufficient for our simulations with Reynolds numbers up to $\SI{1e7}{}$. To this end, we perform DNS with $\mathrm{Re}=\SI{1e7}{}$ for various $n$ and analyze the results for the DJ and DT flows. This involves reproducing the TKE spectra as predicted by the Kraichnan-Batchelor-Leith (KBL) theory \cite{kraichnan_inertial_1967, batchelor_computation_1969, leith_atmospheric_1971}, which describes the energy cascade mechanisms for 2D turbulence. According to this theory, the TKE is proportional to the wavenumber $\kappa^{-3}$ in the inertial range for infinite $\mathrm{Re}$. The inertial range refers to the middle portion of the energy spectrum, between the largest energy-containing eddies and the smallest dissipative scales. In this range, energy is neither injected nor dissipated but transferred progressively.
The TKE spectrum for the DJ and DT simulations is illustrated in Fig.\,\ref{fig:grid_convergence} for several $n$ values at $t=2$, when the turbulence is fully developed.
Our analysis reveals consistency across the largest and most critical scales (small $\kappa$), with all spectra adhering to the expected scaling law. However, the observed dissipation range varies for different $n$. With larger $n$, we can resolve smaller length scales. An ideal simulation would require an $n$ value that ensures that the energy spectrum remains unchanged for higher $n$ values. Nevertheless, as the inertial range is accurately resolved and the dissipation range is merely shifted, we consider a grid resolution with $n=11$ valid for our study's objectives. For a grid resolution with $n=10$, the simulation fails to capture the dissipation range adequately as the energy is elevated in the dissipation range.

\subsection{MPS Compression Efficiency}\label{sec:approx}
\begin{figure}[h]
    \centering
    \includegraphics{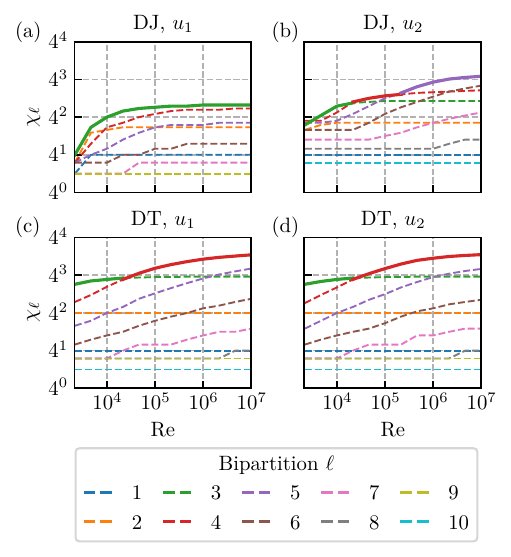}
    \caption{The impact of increasing Reynolds number, $\mathrm{Re}$, on the virtual bond dimensions $\chi_\ell$, such that the MPS representation of the velocity extracted from DNS always has a smaller error than $\epsilon=0.01$. The simulations were conducted on a grid with $n=11$, resulting in 10 bipartitions $\ell$. The maximum bond dimension is $\chi=\max_{\ell}\chi_\ell$. Panels (a) and (b) display $\chi_\ell$ for the $x_1$ and $x_2$ components of the DJ flow, while panels (c) and (d) depict $\chi_\ell$ for the $x_1$ and $x_2$ components of the DT problem. The maximum bond dimension $\chi$ is highlighted by the thick solid lines.}
    \label{fig:chi_Re}
\end{figure}

In Sections \ref{sec:runtime} and \ref{sec:memory}, we showed that the quantum-inspired algorithm with $\chi=64$ offers a runtime and memory advantage for sufficiently large $n$. However, it is still unclear how large $\chi$ needs to be, as its choice directly impacts the accuracy of the QIS. Therefore, we analyze DNS results for various $\mathrm{Re}$ and calculate the required $\chi$ such that the MPSs $\ket{u_i(t, \chi)}$ accurately represent the velocities $\ket{u_i(t)}$ with a maximal error
\begin{equation}
    \epsilon = \frac{\norm{\ket{u_i(t, \chi)} - \ket{u_i(t)}}_2}{\norm{\ket{u_i(t)}}_2}
    \label{eq:epsilon}
\end{equation}
during the simulated time window $t\in[0, 2]$. In this analysis, the maximum bond dimension $\chi$ corresponds to the maximum number of Schmidt values needed to achieve an $\epsilon$-close representation for any bipartition:
\begin{equation}
    \chi = \max_\ell\chi_\ell\mathrm{,}
\end{equation}
where $\chi_\ell$ is the number of Schmidt values for bipartition $\ell$.
Fig.\,\ref{fig:chi_Re} shows $\chi_\ell$ against $\mathrm{Re}$ for the velocity components $u_1$ and $u_2$ of the DJ and DT flow for $\epsilon=0.01$, $n=11$, and a time step of $\Delta t = 0.1/2^{10}$. 
For the DJ flow, the $u_2$ component requires a larger $\chi$ than the $u_1$ component to achieve the same accuracy for similar Reynolds number. This anisotropy is a result of the initial conditions of the DJ flow as discussed in Sec.\,\ref{sec:verification}. On the other hand, the DT flow shows isotropic behavior of $\chi_\ell$ as it is initialized without favoring a spatial direction.

Since we examine the bond dimensions for every bipartition over an extensive range of Reynolds numbers, we reveal the individual behavior of the correlations between two grids containing complementary length scales. Thus, the maximum bond dimension $\chi$ corresponds to $\chi_\ell$ of the dominant bipartition. Fig.\,\ref{fig:chi_Re}(a) demonstrates that this does not need to be the bipartition containing the largest amount of Schmidt values.

By comparing both flows, one can observe that the required $\chi$ is always larger for DT, illustrating its more chaotic nature. Hence, $\chi$ reasonably quantifies the chaotic behavior whereas the Reynolds number only quantifies the level of turbulence for similar flow problems.

Furthermore, it seems like every bond dimension $\chi_\ell$ saturates for large Reynolds numbers. Hence, the maximum bond dimension $\chi$ saturates below its theoretical maximum of $4^5$\footnote{The theoretical maximum of $\chi$ to express an arbitrary state is $d^{\left\lfloor\frac{n}{2}\right\rfloor}$, where $d$ is the dimension of the physical legs $\omega_k$.}. If $\chi$ does not further increase for large Reynolds numbers, we can use the constant saturated value $\chi_\mathrm{sat}(\epsilon) = \chi(\mathrm{Re}\gg1, \epsilon)$ for quantum-inspired simulations of turubulent flows. This leads to a simplification of the computational complexity of QIS from $\mathcal{O}(n\chi^4)$ to $\mathcal{O}(n)$, which is exponentially more efficient than DNS with $\mathcal{O}(4^nn)$. Hence, our findings confirm previous results \cite{gourianov_quantum-inspired_2022}, although we report a different saturated maximum bond dimension $\chi_\mathrm{sat}(\epsilon=0.01)$ of 72 instead of 25. Appendix \ref{sec:discrepancy} provides an explanation for this discrepancy.

\begin{figure}[h]
    \centering
    \includegraphics[width=\textwidth/2]{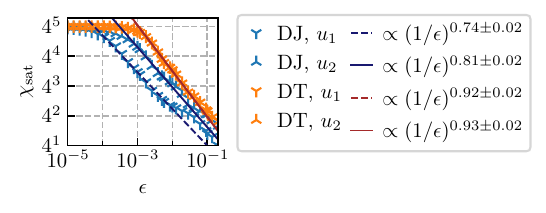}
    \caption{Saturated maximum bond dimension, $\chi_\mathrm{sat}$, depending on the desired representation error $\epsilon$ for $n=11$. The fitted lines highlight the scaling behavior $\chi_\mathrm{sat}=\mathcal{O}(\text{poly}(1/\epsilon))$.}
    \label{fig:chi_sat}
\end{figure}

So far, we have shown that not only the Reynolds number but also the initial conditions influence the required bond dimensions $\chi_\ell(\mathrm{Re},\epsilon)$. To study the impact of the desired maximal error $\epsilon$, we conducted the previous analysis for several $\epsilon$ and observed that $\chi_\mathrm{sat}(\epsilon)$ increases with smaller $\epsilon$ as can be seen in Fig.\,\ref{fig:chi_sat}. 
The less chaotic behavior of the DJ compared to the DT example is reflected in a consistently smaller $\chi_\mathrm{sat}$. For sufficiently small $\epsilon$, $\chi_\mathrm{sat}$ reaches its maximum of $4^5$. For larger $n$, $\chi_\mathrm{sat}$ would continuously grow with smaller $\epsilon$. Our data indicates that $\chi_\mathrm{sat}=\mathcal{O}(\text{poly}(1/\epsilon))$ for the range of $\epsilon$ where $\chi_\mathrm{sat}$ has not reached its maximum. A theoretical explanation for this scaling is given in Sec.\,\ref{sec:derivation}.

\subsection{Theoretical Considerations for Approximating Turbulent Flows with MPSs}\label{sec:derivation}
In this section, we provide a theoretical explanation for the saturation of $\chi$ shown in Fig.\,\ref{fig:chi_Re} and the polynomial scaling of $\chi_\mathrm{sat}$ with $1/\epsilon$ in Fig.\,\ref{fig:chi_sat}.
According to the KBL theory \cite{kraichnan_inertial_1967, batchelor_computation_1969, leith_atmospheric_1971}, 2D turbulence involves two energy cascade mechanisms due to the conservation of kinetic energy and enstrophy in the inertial range. These quantities are conserved only if viscous effects are negligible, which is the case for turbulent flows ($\mathrm{Re}\gg1$). This leads to an inverse energy cascade and a direct enstrophy cascade. Under these assumptions, dimensional analysis shows that the TKE spectrum scales as $E(\kappa)\propto\kappa^{-3}$ in the enstrophy cascade range and as $E(\kappa)\propto\kappa^{-5/3}$ in the inverse energy cascade range. For the case of decaying turbulence, however, the inverse energy cascade does not exist \cite{lesieur_turbulence_2008}, which is why Fig.\,\ref{fig:grid_convergence} only shows the $\kappa^{-3}$ scaling. Since the TKE is directly proportional to the Fourier coefficients of the fluctuating part of the velocity $\hat{u}_i'(\kappa_1, \kappa_2)$, it follows that $E(\kappa)\propto\kappa^{-3}$ leads to $\abs{\hat{u}_i'(\kappa_1, \kappa_2)}\propto\kappa^{-2}$ (see Appendix \ref{sec:derivation_scaling} for details). On the other hand, the mean part of the velocity $\hat{u}_i^\text{mean}(\kappa_1, \kappa_2)$ represents the average flow, which is smoother and contains large-scale features. Consequently, the overall Fourier coefficients of the velocity
\begin{equation}
    \hat{u}_i(\kappa_1, \kappa_2) = \hat{u}_i^\text{mean}(\kappa_1, \kappa_2) + \hat{u}_i'(\kappa_1, \kappa_2)
\end{equation}
are dominated by the velocity fluctuations at high $\kappa$. Thus, the Fourier coefficients decay as
\begin{equation}
    \abs{\hat{u}_i(\kappa_1, \kappa_2)} \leq C\kappa^{-2}
\end{equation}
for wave numbers of the inertial range. Here, $C$ is a constant. Using this upper bound for the Fourier coefficients, the theoretical analysis of Ref.\,\cite{jobst_efficient_2023} shows that the maximum bond dimension scales with $\chi_\mathrm{sat}=\mathcal{O}(\text{poly}(1/\epsilon))$. This explains the observed polynomial scaling in Fig.\,\ref{fig:chi_sat}. The derivation and more details are given in Appendix \ref{sec:derivation_scaling}. 

For finite Reynolds numbers, the TKE spectrum does not necessarily follow the $\kappa^{-3}$ power law. However, as the Reynolds number increases, the flow approaches an inviscid state, and the TKE spectrum correspondingly approaches the $\kappa^{-3}$ power law. Accordingly, the maximum bond dimensions in Fig.\,\ref{fig:chi_Re} approach their saturated values in the limit of high Reynolds numbers. 

Above the so-called Kolmogorov wave number $\kappa_\mathrm{d}$, the dissipation range begins as viscous effects cause energy dissipation. In the dissipation range ($\kappa>\kappa_\mathrm{d}$), the TKE spectrum decays rapidly (possibly exponentially \cite{lesieur_turbulence_2008}). Thus, we can even assume an exponential decay of the Fourier coefficients $\abs{\hat{u}_i(\kappa_1, \kappa_2)}$ for $\kappa\gg1$. For smaller Reynolds numbers, $\kappa_\mathrm{d}$ shifts towards smaller wave numbers. Consequently, for finite Reynolds numbers, the maximum bond dimension $\chi$ will be below $\chi_\mathrm{sat}$ since the Fourier coefficients above $\kappa_\mathrm{d}$ are negligible.

Considering the relation $\chi=\mathcal{O}(\text{poly}(1/\epsilon))$, the overall complexity of the quantum-inspired algorithm can be written as $\mathcal{O}(n\,\text{poly}(1/\epsilon))$.
\subsection{Quantum-Inspired Simulations for high Reynolds Numbers}
\begin{figure}[h]
    \centering
    \includegraphics{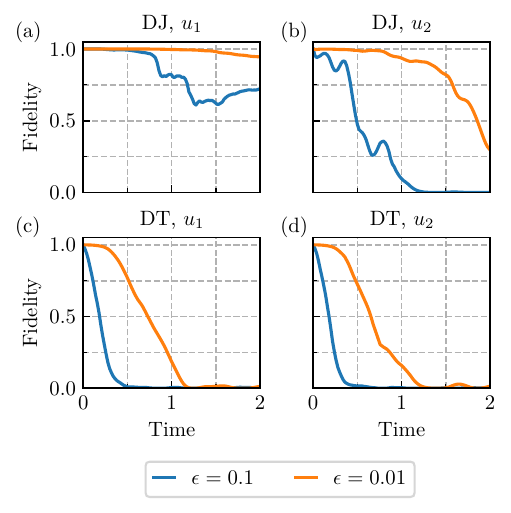}
    \caption{Fidelities according to Eq.\eqref{eq:fidelity} between DNS and QIS with $\mathrm{Re}=\SI{1e7}{}$ for errors $\epsilon$ corresponding to respective $\chi$ values (cf. Table\,\ref{tab:QIS_params}). Panels (a) and (b) show the fidelities for the $u_1$ and $u_2$ components of DJ simulations, while panels (c) and (d) illustrate the fidelities for the $u_1$ and $u_2$ components of the DT example.}
    \label{fig:fidelity}
\end{figure}

In Sec.\,\ref{sec:verification}, we have verified that the QIS yields similar results as the DNS for DJ flow with $\mathrm{Re}=\SI{2e5}{}$. However, $\chi$ is not saturated at this Reynolds number (cf. Fig.\,\ref{fig:chi_Re}), which is why we also ran QIS with $\mathrm{Re}=\SI{1e7}{}$. We simulated both flow problems with a grid resolution corresponding to $n=11$. Fig.\,\ref{fig:fidelity} shows the fidelities (cf. Eq.\,\eqref{eq:fidelity}) for QISs with maximum bond dimension $\chi$ corresponding to errors of $\epsilon=0.1$ and $\epsilon=0.01$ (cf. Table\,\ref{tab:QIS_params}). 
\begin{table}[h]
    \centering
    \begin{tabular}{c c c c c}
        \toprule
         & $\epsilon=0.1$ & $\epsilon=0.01$\\
        \midrule
        DJ & $\chi_\mathrm{sat}=17$ & $\chi_\mathrm{sat}=72$\\
        DT & $\chi_\mathrm{sat}=20$ & $\chi_\mathrm{sat}=137$\\
        \bottomrule
    \end{tabular}
    \caption{Simulation parameters for the QIS with different error $\epsilon$ corresponding to a maximum bond dimension $\chi$. Here, $\chi_\mathrm{sat}$ is extracted from our analysis in Fig.\,\ref{fig:chi_sat}.}
    \label{tab:QIS_params}
\end{table}
\noindent For the DJ flow, the fidelity behaves similarly as in Fig.\,\ref{fig:fidelities_low_Re}, showcasing the anisotropy of the flow. In contrast, both components of the DT simulations exhibit similar trends. The impact of $\epsilon$ is evident, as smaller $\epsilon$ values enhance fidelity, indicating that QIS can achieve comparable accuracy to DNS with an adequately selected $\epsilon$ based on $\chi$. However, our data shows that even an error of $\epsilon=0.01$ is insufficient to achieve practical results at $t=2$. It should be noted that other error sources exist beyond the compression with $\chi$. The DMRG-like optimization sweeps end as soon as the relative change compared to the previous solution is below $\SI{1e-5}{}$. Similarly, the CG solver ends if the residual is smaller than $\SI{1e-5}{}$ or after 100 iterations. Decreasing these thresholds and increasing the number of CG iterations would increase the fidelity.

\begin{figure}[h]
    \centering
    \includegraphics{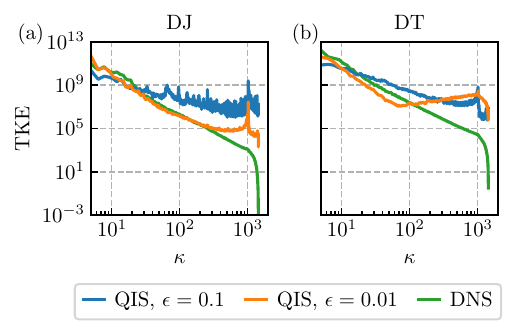}
    \caption{TKE spectra for (a) the DJ flow and (b) the DT flow with $\mathrm{Re} = \SI{1e7}{}$ at $t=2$. Errors $\epsilon$ correspond to $\chi$ values according to Table\,\ref{tab:QIS_params}.}
    \label{fig:E_k_QIS}
\end{figure}

In addition to the fidelity, we computed the TKE spectra of both flows at $t=2$ in Fig.\,\ref{fig:E_k_QIS}. 
\noindent As previously observed, a smaller $\epsilon$ for the DJ flow leads to better agreement in the inertial range but discrepancies emerge at higher $\kappa$. These discrepancies are particularly significant for the DT flow, as the TKE spectrum for $\epsilon=0.01$ looks distorted from the DNS results. While small errors at high $\kappa$ are expected, the quantum-inspired CFD algorithm fails to accurately approximate the TKE spectrum of the DT flow under this extreme turbulent configuration.

\section{Conclusion and Outlook}\label{sec:discussion}
In this work, we have examined the quantum-inspired algorithm for simulating 2D turbulent flows, as introduced by \cite{gourianov_quantum-inspired_2022, gourianov_exploiting_2022}, and have extended it with a higher order and more accurate RK4 time-stepping scheme to simulate turbulent flows with Reynolds numbers up to $\SI{1e7}{}$. Beyond that, we have leveraged GPU parallelization for tensor operations, achieving up to 12.1-fold speedup, which is crucial for facilitating the practical applications of this class of algorithms. Additionally, the algorithm's runtime and memory consumption have been analyzed and compared with the DNS results, identifying regimes where QIS becomes advantageous. Moreover, this study also evaluates the efficiency of the MPS encoding in capturing the velocity field for simulations with high $\mathrm{Re}$, concluding that the maximum bond dimension $\chi(\mathrm{Re}, \epsilon)$ saturates for high $\mathrm{Re}$ at $\chi_\mathrm{sat}(\epsilon)$ below its theoretical maximum. By setting $\chi=\chi_\mathrm{sat}(\epsilon)$, the algorithm theoretically offers an exponential complexity advantage. Thereby, it is essential that the system size $n=\log_2N$ for an $N\times N$ grid is large enough to resolve all flow characteristics. To verify that an appropriate grid resolution has been chosen, a grid convergence analysis has been carried out showing that $n=11$ is sufficient to display a complete TKE spectrum, though capturing the smallest flow scales would require a larger $n$. Given that the TKE spectrum follows a characteristic power law for turbulent flows, we infer a similar power law for the Fourier coefficients of the velocity fields. This allows us to deduce $\chi_\mathrm{sat}=\mathcal{O}(\text{poly}(1/\epsilon))$, which is consistent with our data and explains the saturation of $\chi$ for large Reynolds numbers. This scaling law, as a novel theoretical justification of \textit{quantics} MPS or QTT encoding, should be valid for a wide range of fluid simulations, as the characteristic energy distribution directly results from conservation laws.

Interestingly, the quantum-inspired algorithm gives good results for moderate Reynolds numbers ($\mathrm{Re}=\SI{2e5}{}$), but for $\mathrm{Re}=\SI{1e7}{}$, a quick decline in fidelity is observed. In addition, TKE spectrum in Fig.\,\ref{fig:E_k_QIS} shows an unusual increase of energy at small length scales. One could improve the algorithm's accuracy by increasing the number of CG iterations or DMRG-like optimization sweeps, which would result in very long runtimes. Creating necessary MPOs and solving four optimization problems per RK4 time step results in an excessive overhead for the algorithm. One approach to reducing runtime is to further distribute these optimization tasks across multiple GPUs. Alternatively, one could employ other TN algorithms with less overhead. Due to the periodicity of the boundary conditions, the DNS scheme consists of repeated FFTs which are extremely efficient. A promising approach is to combine the MPS/MPO encoding with the Quantum Fourier Transform (QFT). Recently, it has been shown that the core routine of QFT can be written as an MPO with low maximum bond dimension \cite{chen_quantum_2023}, thus, applying the QFT-MPO for such problems can be lead to more computational efficiency. Since we have shown that the velocities of turbulent flows can be well approximated with MPSs, CFD simulations with periodic boundary conditions are the ideal use case for QFT as TN algorithm. Moreover, a major part of TN algorithms are repeated SVDs to keep the maximum bond dimension of MPSs and MPOs low. The tensor cross interpolation (TCI) seems to be a promising alternative, as it is exponentially more efficient than SVDs while yielding decompositions of similar quality \cite{ritter_quantics_2024}.  

In addition, future research will investigate the application of quantum-inspired TN algorithms for simulating 3D flows. If these methods prove advantageous for 3D fluid simulations, they would have a significant impact, benefiting numerous real-world applications. Although 3D turbulence is inherently different from 2D turbulence, the TKE spectrum of 3D flows also follows a characteristic power law ($E(\kappa)\propto\kappa^{-5/3}$) in the inertial range with an exponential drop in the dissipation range \cite{pope_turbulent_2000}. The less steep decay of the TKE spectrum with $\kappa$ leads to a less steep decay of the Fourier coefficients of the velocity components. Thus, the required $\chi$ to achieve an $\epsilon$-close MPS representation of the velocity component might be higher than for the 2D case. Nevertheless, due to the power law of the kinetic energy distribution, the decay of Fourier coefficients is guaranteed, validating the potential of an efficient MPS approximation. The analysis of Gourianov et al. \cite{gourianov_quantum-inspired_2022} included the simulation of a 3D Taylor-Green vortex flow. Their analysis did not show a saturation of $\chi$ below its maximum for large $\mathrm{Re}$. However, our results show that the saturation of $\chi$ could arise for even higher $\mathrm{Re}$, larger $n$, or different $\epsilon$, which needs to be checked by future simulations.

Finally, the insights of this work benefit other areas of research as well. As TN methods originate from simulating quantum systems, this work also highlights the potential to speed up such simulations by parallelizing huge contractions using GPUs. Moreover, the algorithm itself is not limited to solving the Navier-Stokes equations. It can be applied to solve other partial differential equations as well \cite{ye_quantum-inspired_2022}. In a different vein, quantum-inspired algorithms are of particular interest for quantum algorithms. The omnipresent data loading problem of quantum computers might be bypassed by using quantum circuits based on MPSs \cite{rudolph_decomposition_2022, jobst_efficient_2023}. Unfortunately, the Navier-Stokes equations' nonlinearity is a significant obstacle for quantum algorithms, as quantum operations are inherently linear. Variational quantum algorithms do not have this limitation, as one can build quantum circuits representing a nonlinear cost function to solve nonlinear problems \cite{lubasch_variational_2020}. This approach has been suggested to solve the Navier-Stokes equations similarly to the presented quantum-inspired CFD algorithm \cite{jaksch_variational_2022}. Hence, we hope that our work contributes to and advances the fields of TN algorithms, fluid dynamics, and quantum algorithms.

\begin{acknowledgments}
We acknowledge the support of the BMW Group and thank Carlos Riofrío for his valuable discussions. We also extend our appreciation to Dmitry Lyakh, Yang Gao, Satya Varadhan, and Jin-Sung Kim from the NVIDIA Quantum team. Additionally, we thank Jeremy Melvin for his valuable insights into turbulence modeling.
\end{acknowledgments}

\bibliographystyle{ieeetr}
\bibliography{references,manual_bib}

\onecolumngrid
\appendix
\newpage
\section{Calculating Derivatives with MPOs}\label{sec:diff_mpos}
Differential MPOs are an essential building block of TN algorithms for solving differential equations. Here, we provide the tensors for the central finite difference operator of second order in $x$-direction $\hat{D}_1$ mentioned in Sec.\,\ref{sec:diff_mpos_theory}. For a $2^n\times2^n$ grid, the operator is defined as
\begin{equation}
    \hat{D}_1 = \sum_{\{\alpha_\ell\}=1}^{\leq\chi}L_{\alpha_{-1}}\left(\bigotimes_{m=0}^{n-1}C_{\alpha_m\alpha_{m+1}}^{\omega_m\omega'_m}\right)R_{\alpha_{n}}\mathrm{.}
\end{equation}
Here, $\{\alpha_\ell\}$ is the set of virtual bonds and $\{\omega_m\}$ (and $\{\omega'_m\}$) is the set of physical bonds of the MPO. The left, right, and central tensors $L_{\alpha_{-1}}$, $R_{\alpha_{n}}$, and $C_{\alpha_m\alpha_{m+1}}^{\omega_m\omega'_m}$ correspond to the tensors in Eq.\,\eqref{eq:mpo_d1}. In order to calculate the derivative according to 
\begin{equation}
    \frac{\partial U_i^{xy}}{\partial x} = \frac{U_i^{(x+1)y}}{2\Delta x} - \frac{U_i^{(x-1)y}}{2\Delta x}\text{.}
    \label{eq:fdm2}
\end{equation}
the edge tensors look like:
\begin{equation}
    L_{\alpha_{-1}} = 
    \begin{pmatrix}
        1&1&1
    \end{pmatrix}
    \quad
    R_{\alpha_{n}} = \frac{1}{2\Delta x}
    \begin{pmatrix}
        0\\
        1\\
        -1
    \end{pmatrix}
\end{equation}
The central node $C_{\alpha_m\alpha_{m+1}}^{\omega_m\omega'_m}$ is a rank-4 tensor where the virtual bonds $\alpha_m$ have dimension 3 and the physical bonds $\omega_m$ have dimension 4 ($2^k$ for a $k$-dimensional grid). All values of the tensor are set to $0$ except the parameters mapping the binary addition/subtraction logic. The mapping considers three cases given by $\alpha_{m+1}$ connected to the tensor of the less significant bits. If $\alpha_{m+1}=0$, the output index $\omega'_m$ should be identical to $\omega_m$. In case $\alpha_{m+1}=1$, we add 1 to the $x$-bit such that $\omega'_m=(x_m+1 y_m)_2$ and $\omega_m=(x_my_m)_2$. In case $\alpha_{m+1}=2$, we subtract 1 of the $x$-bit such that the output index is $\omega'_m=(x_m-1 y_m)_2$. If the result cannot be represented by the bit $x_m$, we need to consider a carry in the form of setting $\alpha_{m}=1$ or $\alpha_{m}=2$ in order to add or subtract 1 from the next significant bit. All resulting cases and resulting tensor values are displayed in Table\,\ref{table:diff_mpo}.
\begin{table}[H]
\centering
\begin{tabular}{c c c c }
\toprule
                        & \qquad$\omega_m\rightarrow\omega'_m$\qquad & \qquad carry \qquad & \qquad$C_{\alpha_m\alpha_{m+1}}^{\omega_m\omega'_m}$\qquad \\ 
\midrule
                        & \qquad$(00)_2\rightarrow(00)_2$ & \qquad$\alpha_{m}=0$ & \qquad$C_{00}^{00}=1$ \\  
$\alpha_{m+1}=0$        & \qquad$(01)_2\rightarrow(01)_2$ & \qquad$\alpha_{m}=0$ & \qquad$C_{00}^{11}=1$ \\
($x_m\pm0$)             & \qquad$(10)_2\rightarrow(10)_2$ & \qquad$\alpha_{m}=0$ & \qquad$C_{00}^{22}=1$ \\
                        & \qquad$(11)_2\rightarrow(11)_2$ & \qquad$\alpha_{m}=0$ & \qquad$C_{00}^{33}=1$ \\
\midrule
                        & \qquad$(00)_2\rightarrow(10)_2$ & \qquad$\alpha_{m}=0$ & \qquad$C_{01}^{02}=1$ \\
$\alpha_{m+1}=1$        & \qquad$(01)_2\rightarrow(11)_2$ & \qquad$\alpha_{m}=0$ & \qquad$C_{01}^{13}=1$ \\
($x_m+1$)               & \qquad$(10)_2\rightarrow(00)_2$ & \qquad$\alpha_{m}=1$ & \qquad$C_{11}^{20}=1$ \\
                        & \qquad$(11)_2\rightarrow(01)_2$ & \qquad$\alpha_{m}=1$ & \qquad$C_{11}^{31}=1$ \\
\midrule
                        & \qquad$(00)_2\rightarrow(10)_2$ & \qquad$\alpha_{m}=2$ & \qquad$C_{22}^{02}=1$ \\
$\alpha_{m+1}=2$        & \qquad$(01)_2\rightarrow(11)_2$ & \qquad$\alpha_{m}=2$ & \qquad$C_{22}^{13}=1$ \\
($x_m-1$)               & \qquad$(10)_2\rightarrow(00)_2$ & \qquad$\alpha_{m}=0$ & \qquad$C_{02}^{20}=1$ \\
                        & \qquad$(11)_2\rightarrow(01)_2$ & \qquad$\alpha_{m}=0$ & \qquad$C_{02}^{31}=1$ \\
\bottomrule
\end{tabular}
\caption{All tensor values of the central node $C_{\alpha_m\alpha_{m+1}}^{\omega_m\omega'_m}$ to build a differential MPO to calculate central finite differences of second order.}
\label{table:diff_mpo}
\end{table}

\section{Iterative Optimization Scheme for Solving the Variational Form of the Incompressible Navier Stokes Equations}\label{sec:cost}
Here, we derive the cost function for the DMRG-like algorithm and explain how it is optimized. A similar derivation can be found in \cite{gourianov_exploiting_2022}; however, for enhanced clarity, we offer a detailed derivation employing our notation. We start with the Navier-Stokes equations written with MPSs and MPOs:
\begin{equation}
    \hat{D}_1\ket{u_1} + \hat{D}_2\ket{u_2} = 0 
    \label{eq:MPS_continuity_appendix}
\end{equation}
\begin{equation}
    \frac{d}{dt}\ket{u_i} = \sum_{j=1}^2 \underbrace{-\frac{1}{2}\left(\ket{u_j}\odot \hat{D}_j\ket{u_i} + \hat{D}_j\left(\ket{u_j}\odot\ket{u_i}\right)\right)}_\text{convection} + \underbrace{\frac{1}{\text{Re}}\hat{D}_j^2\ket{u_i}}_\text{diffusion} - \hat{D}_i \ket{p} \quad\text{for }i\in\{1, 2\}\text{.}
    \label{eq:MPS_momentum_appendix}
\end{equation}
Here, we have used the skew-symmetric representation of the convection term for better numerical stability \cite{canuto_spectral_2007, vuorinen_dnslab_2016}.
Now, we systematically rearrange the equations by isolating all terms on the left-hand side. Subsequently, we compute the squared norm of the terms and sum them up to obtain the cost function
\begin{align}
\begin{split}
    \Omega\left(\ket{v_1}, \ket{v_2}\right) = &\,\mu \norm{\hat{D}_1\ket{v_1} + \hat{D}_2\ket{v_2}}^2 \\
    & + \norm{\sum_{i=1}^2\bm{e}_i \left[\frac{\partial}{\partial t}\ket{v_i} + \sum_{j=1}^2 \frac{1}{2}\left(\ket{v_j}\odot \hat{D}_j\ket{v_i} + \hat{D}_j\left(\ket{v_j}\odot\ket{v_i}\right)\right) - \frac{1}{\text{Re}}\hat{D}_j^2\ket{v_i}\right]}^2\text{.}
    \label{eq:cost_function}
\end{split}
\end{align}

Here, $\bm{e}_i$ is the unit vector in $i$ direction and the velocity field $\left(\ket{u_1}, \ket{u_2}\right)$ satisfying the Navier-Stokes yields $\Omega\left(\ket{u_1}, \ket{u_2}\right)=0$. Thus, we receive the solution by minimizing $\Omega$:
\begin{equation}
    \left(\ket{u_1}, \ket{u_2}\right) = \underset{\ket{v_1}, \ket{v_2}}{\arg\min}\ \Omega\left(\ket{v_1}, \ket{v_2}\right)\text{.}
\end{equation}
In Eq.\,\eqref{eq:cost_function}, the pressure term disappears because the incompressibility condition is enforced by the penalty factor $\mu$. For all our simulations, we used $\mu=\SI{25e4}{}$ as suggested in \cite{gourianov_exploiting_2022}.
The time derivative is approximated by the simple difference quotient
\begin{equation}
    \frac{\partial}{\partial t}\ket{v_i} \approx \frac{\ket{v_i^{s+1}} - \ket{v_i^{s}}}{h}\mathrm{,}
    \label{eq:euler}
\end{equation}
where $s+1$ marks the next and $s$ the previous time step. The relative time step is given by $\Delta t = t_s-t_{s-1}$. If we use the approximation in Eq.\,\eqref{eq:euler} to evaluate the cost function in Eq.\,\eqref{eq:cost_function}, we end up using the explicit Euler method (RK1) for time stepping. To achieve more accurate solutions, one can employ higher order RK methods. For RK2 time stepping, we additionally evaluate the velocity field at a midpoint $s+0.5$. Therefore, we rewrite the cost function to consider velocities of different time steps as input:
\begin{align}
\begin{split}
    \Theta&\left(\ket{v_1}, \ket{v_2}, \ket{a_1}, \ket{a_2}, \ket{b_1}, \ket{b_2}, h\right) = \mu \norm{\mathrm{D_1}\ket{v_1} + \mathrm{D_2}\ket{v_2}}^2 \\
    &+ \norm{\sum_{i=1}^2\bm{e}_i \left[\frac{\ket{v_i} - \ket{a_i}}{h} + \sum_{j=1}^2 \frac{1}{2}\left(\ket{b_j}\odot \hat{D}_j\ket{b_i} + \hat{D}_j\left(\ket{b_j}\odot\ket{b_i}\right)\right) - \frac{1}{\text{Re}}\hat{D}_j^2\ket{b_i}\right]}^2\text{.}
    \label{eq:Theta}
\end{split}
\end{align}
Thus, the velocity field at the midpoint step in time is calculated as
\begin{equation}
    \left(\ket{u_1^{s+0.5}}, \ket{u_2^{s+0.5}}\right) = \underset{\ket{v_1}, \ket{v_2}}{\arg\min}\ \Theta\left(\ket{v_1}, \ket{v_2}, \ket{u_1^s}, \ket{u_2^s}, \ket{u_1^s}, \ket{u_2^s}, \Delta t/2\right)\text{.}
    \label{eq:midpoint_calc}
\end{equation}
The velocity after a full time step calculated as
\begin{equation}
    \left(\ket{u_1^{s+1}}, \ket{u_2^{s+1}}\right) = \underset{\ket{v_1}, \ket{v_2}}{\arg\min}\ \Theta\left(\ket{v_1}, \ket{v_2}, \ket{u_1^s}, \ket{u_2^s}, \ket{u_1^{s+0.5}}, \ket{u_2^{s+0.5}}, \Delta t\right)\text{.}
    \label{eq:full_calc}
\end{equation}
To employ \textbf{RK4 time stepping}, as we have done in the paper, we have to perform four minimizations. Therefore, we split up the explicit RK4 equation into four Euler steps:
\begin{align}
    \ket{u_i^{s+1}} &= \ket{u_i^{s}} + \frac{\Delta t}{6}\left(\ket{k_1}+2\ket{k_2}+2\ket{k_3}+\ket{k_4}\right)\\
                    &= \underbrace{\frac{\ket{u_i^{s}}}{4} + \frac{\Delta t}{6}\ket{k_1}}_{=\ket{U1_i}} + \underbrace{\frac{\ket{u_i^{s}}}{4} + \frac{\Delta t}{3}\ket{k_2}}_{=\ket{U2_i}} + \underbrace{\frac{\ket{u_i^{s}}}{4} + \frac{\Delta t}{3}\ket{k_3}}_{=\ket{U3_i}} + \underbrace{\frac{\ket{u_i^{s}}}{4} + \frac{\Delta t}{6}\ket{k_4}}_{=\ket{U4_i}}\mathrm{.}\label{eq:4euler}
\end{align}
Here, the RK4 gradients $\ket{k_{j}}_{j\in\{1, 2, 3, 4\}}$ are calculated implicitly via $\ket{b_1}$ and $\ket{b_2}$ in Eq.\,\eqref{eq:Theta}. Table \ref{tab:RK4} shows the definition of $\ket{b_1}$ and $\ket{b_2}$ according to \cite{ferziger_computational_2020}. From Eq.\,\ref{eq:4euler}, we can see that the RK4 gradients can be calculated from the respective intermediate Euler steps $\ket{U1_i}$, $\ket{U2_i}$, and $\ket{U3_i}$.
\begin{table}[H]
    \centering
    \begin{tabular}{l l l l}
        \toprule
        First RK step: &\quad&$\ket{b_1} = \ket{u_1^s}$ & $\ket{b_2} = \ket{u_2^s}$\\[15pt]
        Second RK step: &\quad&$\ket{b_1} = \ket{u_1^s} + \frac{\Delta t}{2}\ket{k_{1,1}} = 3\ket{U1_1}+\frac{1}{4}\ket{u_1^s}$ & $\ket{b_2} = \ket{u_2^s} + \frac{\Delta t}{2}\ket{k_{1,2}} = 3\ket{U1_2}+\frac{1}{4}\ket{u_2^s}$\\[15pt]
        Third RK step: &\quad&$\ket{b_1} = \ket{u_1^s} + \frac{\Delta t}{2}\ket{k_{2,1}} = \frac{3}{2}\ket{U2_1}+\frac{5}{8}\ket{u_1^s}$ & $\ket{b_2} = \ket{u_2^s} + \frac{\Delta t}{2}\ket{k_{2,2}} = \frac{3}{2}\ket{U2_2}+\frac{5}{8}\ket{u_2^s}$\\[15pt]
        Fourth RK step: &\quad&$\ket{b_1} = \ket{u_1^s} + \Delta t\ket{k_{3,1}} = 3\ket{U3_1}+\frac{1}{4}\ket{u_1^s}$ & $\ket{b_2} = \ket{u_2^s} + \Delta t\ket{k_{3,2}} = 3\ket{U3_2}+\frac{1}{4}\ket{u_2^s}$\\
        \bottomrule
    \end{tabular}
    \caption{Definition of $\ket{b_1}$ and $\ket{b_2}$ for the RK4 scheme according to \cite{ferziger_computational_2020}.}
    \label{tab:RK4}
\end{table}
Consequently, the Euler steps are computed as
\begin{align}
    \left(\ket{U1_1}, \ket{U1_2}\right) &= \underset{\ket{v_1}, \ket{v_2}}{\arg\min}\ \Theta\left(\ket{v_1}, \ket{v_2}, \frac{\ket{u_1^s}}{4}, \frac{\ket{u_2^s}}{4}, \ket{u_1^s}, \ket{u_2^s}, \frac{\Delta t}{6}\right)\label{eq:U1}\\
    \left(\ket{U2_1}, \ket{U2_2}\right) &= \underset{\ket{v_1}, \ket{v_2}}{\arg\min}\ \Theta\left(\ket{v_1}, \ket{v_2}, \frac{\ket{u_1^s}}{4}, \frac{\ket{u_2^s}}{4}, 3\ket{U1_1}+\frac{\ket{u_1^s}}{4}, 3\ket{U1_2}+\frac{\ket{u_2^s}}{4}, \frac{\Delta t}{3}\right)\label{eq:U2}\\
    \left(\ket{U3_1}, \ket{U3_2}\right) &= \underset{\ket{v_1}, \ket{v_2}}{\arg\min}\ \Theta\left(\ket{v_1}, \ket{v_2}, \frac{\ket{u_1^s}}{4}, \frac{\ket{u_2^s}}{4}, \frac{3}{2}\ket{U2_1}+\frac{5}{8}\ket{u_1^s}, \frac{3}{2}\ket{U2_2}+\frac{5}{8}\ket{u_2^s}, \frac{\Delta t}{3}\right)\label{eq:U3}\\
    \left(\ket{U4_1}, \ket{U4_2}\right) &= \underset{\ket{v_1}, \ket{v_2}}{\arg\min}\ \Theta\left(\ket{v_1}, \ket{v_2}, \frac{\ket{u_1^s}}{4}, \frac{\ket{u_2^s}}{4}, 3\ket{U3_1}+\frac{\ket{u_1^s}}{4}, 3\ket{U3_2}+\frac{\ket{u_2^s}}{4}, \frac{\Delta t}{6}\right)\mathrm{.}\label{eq:U4}
\end{align}
The minimizations in Eqs.\,\eqref{eq:midpoint_calc}, \eqref{eq:full_calc}, or Eqs.\,\eqref{eq:U1}, \eqref{eq:U2}, \eqref{eq:U3}, \eqref{eq:U4} are done by iteratively updating each tensor of the MPSs, such as in the DMRG method. Therefore, we transform the velocity MPS of interest
\begin{equation}
    \ket{v_i} = \sum_{\{\omega_k\}=0}^3 \sum_{\{\alpha_\ell\}=1}^{\leq\chi} V^{\omega_0}_{i, \alpha_0} V^{\omega_1}_{i, \alpha_0\alpha_1} \ldots V^{\omega_{n-1}}_{i, \alpha_{n-2}} \ket{\omega_0\omega_1\ldots\omega_{n-1}}
\end{equation}
in mixed-canonical form, such that the $d$-th tensor is the canonical center of the MPS
\begin{equation}
    \ket{v_i} = \sum_{\{\omega_k\}=0}^3 \sum_{\{\alpha_\ell\}=1}^{\leq\chi} L^{\omega_0}_{i, \alpha_0}\ldots L^{\omega_{d-1}}_{i, \alpha_{d-2}\alpha_{d-1}} C^{\omega_d}_{i, \alpha_{d-1}\alpha_{d}} R^{\omega_{d+1}}_{i, \alpha_{d}\alpha_{d+1}}\ldots R^{\omega_{n-1}}_{i, \alpha_{n-1}} \ket{\omega_0\dots\omega_{d-1}\omega_{d}\omega_{d+1}\dots\omega_{n-1}} \mathrm{.}
\end{equation}
Here, $L^{\omega_{k}}_{i, \alpha_{k-1}\alpha_{k}}$ and $R^{i, \omega_{k}}_{\alpha_{k-1}\alpha_{k}}$ are left and right isometries, respectively. $C^{\omega_d}_{i, \alpha_{d-1}\alpha_{d}}$ represents the canonical center and describes the whole MPS $\ket{v_i}$ with new basis states
\begin{equation}
    \ket{\alpha_{d-1}^i} = \sum_{\{\omega_1\dots\omega_{d-1}\}=0}^3 \sum_{\{\alpha_0\dots\alpha_{d-2}\}=1}^{\leq\chi} L^{\omega_0}_{i, \alpha_0}\ldots L^{\omega_{d-1}}_{i, \alpha_{d-2}\alpha_{d-1}} \ket{\omega_0\ldots\omega_{d-1}}\mathrm{,}
\end{equation}
\begin{equation}
    \ket{\alpha_d^i} = \sum_{\{\omega_{d+1}\dots\omega_n\}=0}^3\sum_{\{\alpha_{d+1}\dots\alpha_{n-1}\}=1}^{\leq\chi} R^{\omega_{d+1}}_{i, \alpha_{d}\alpha_{d+1}}\ldots R^{\omega_{n-1}}_{i, \alpha_{n-1}} \ket{\omega_{d+1}\ldots\omega_{n-1}}
\end{equation}
and $\ket{\omega_d}$.
Thus, we can write
\begin{equation}
    \ket{v_i} = \sum_{\alpha_{d-1}=1}^{\leq\chi}\sum_{\omega_d=0}^3\sum_{\alpha_{d}=1}^{\leq\chi} C^{\omega_d}_{i, \alpha_{d-1}\alpha_{d}} \ket{\alpha_{d-1}^i\omega_d\alpha_d^i}\mathrm{.}
    \label{eq:mixed_canonical}
\end{equation}
The tensors $L^{\omega_{k}}_{i, \alpha_{k-1}\alpha_{k}}$, $R^{i, \omega_{k}}_{\alpha_{k-1}\alpha_{k}}$, and $C^{\omega_d}_{i, \alpha_{d-1}\alpha_{d}}$ should not be confused with those from the previous Sec.\,\ref{sec:diff_mpos}. To find the minimum of $\Theta$, we calculate the derivative
\begin{equation}
    \frac{\partial\Theta}{\partial C^{\omega_d}_{i, \alpha_{d-1}\alpha_{d}}} = \bra{\alpha_{d-1}^i\omega_d\alpha_d^i} \left[\left(\sum_{j=1}^22\mu\hat{D}_i^\dagger\hat{D}_j\ket{v_j}\right) + \frac{2}{\Delta t^2}\ket{v_i} - \frac{2}{\Delta t^2}\ket{a_i} + \frac{2}{\Delta t}\ket{B_i}\right]
\end{equation}
with
\begin{equation}
    \ket{B_i} = \sum_{j=1}^2 \frac{1}{2}\left(\ket{b_j}\odot \hat{D}_j\ket{b_i} + \hat{D}_j\left(\ket{b_j}\odot\ket{b_i}\right)\right) - \frac{1}{\text{Re}}\hat{D}_j^2\ket{b_i}
\end{equation}
and set it equal to zero
\begin{equation}
    \frac{\partial\Theta}{\partial C^{\omega_d}_{i, \alpha_{d-1}\alpha_{d}}} = 0\mathrm{.}
\end{equation}
This gives us two coupled systems of equations as we consider the derivative for every tensor value of $C^{\omega_d}_{i, \alpha_{d-1}\alpha_{d}}$ for $i\in\{1, 2\}$:
\begin{align}
    C^{\omega_d}_{1, \alpha_{d-1}\alpha_{d}} - \mu\Delta t^2\sum_{j=1}^2\sum_{\alpha_{d-1}'\omega_d'\alpha_d'} C^{\omega_d'}_{j, \alpha_{d-1}'\alpha_{d}'}\bra{\alpha_{d-1}^1\omega_d\alpha_d^1}\hat{D}_1\hat{D}_j\ket{\alpha^{\prime j}_{d-1}\omega_d'\alpha_d^{\prime j}} = \braket{\alpha_{d-1}^1\omega_d\alpha_d^1}{a_1} - \Delta t \braket{\alpha_{d-1}^1\omega_d\alpha_d^1}{B_1} \label{eq:first}\\
    C^{\omega_d}_{2, \alpha_{d-1}\alpha_{d}} - \mu\Delta t^2\sum_{j=1}^2\sum_{\alpha_{d-1}'\omega_d'\alpha_d'} C^{\omega_d'}_{j, \alpha_{d-1}'\alpha_{d}'}\bra{\alpha_{d-1}^2\omega_d\alpha_d^2}\hat{D}_2\hat{D}_j\ket{\alpha^{\prime j}_{d-1}\omega_d'\alpha_d^{\prime j}} = \braket{\alpha_{d-1}^2\omega_d\alpha_d^2}{a_2} - \Delta t \braket{\alpha_{d-1}^2\omega_d\alpha_d^2}{B_2} \label{eq:second}
\end{align}
Here, we used the relation $\hat{D}_i^\dagger=-\hat{D}_i$. By combining Eqs.\,\eqref{eq:first} and \eqref{eq:second}, we can build a global linear system

\begin{multline}
    \left[\mathbb{1}-\mu\Delta t^2
    \begin{pNiceArray}{ccc|ccc}
        \ddots & \vdots & \iddots & \ddots & \vdots & \iddots \\
        \ldots & \bra{\alpha_{d-1}^1\omega_d\alpha_d^1}\hat{D}_1\hat{D}_1\ket{\alpha^{\prime 1}_{d-1}\omega_d'\alpha_d^{\prime 1}} & \ldots & \ldots & \bra{\alpha_{d-1}^1\omega_d\alpha_d^1}\hat{D}_1\hat{D}_2\ket{\alpha^{\prime 2}_{d-1}\omega_d'\alpha_d^{\prime 2}} & \ldots \\
        \iddots & \vdots & \ddots & \iddots & \vdots & \ddots \\
        \hline
        \ddots & \vdots & \iddots & \ddots & \vdots & \iddots \\
        \ldots & \bra{\alpha_{d-1}^2\omega_d\alpha_d^2}\hat{D}_2\hat{D}_1\ket{\alpha^{\prime 1}_{d-1}\omega_d'\alpha_d^{\prime 1}} & \ldots & \ldots & \bra{\alpha_{d-1}^2\omega_d\alpha_d^2}\hat{D}_2\hat{D}_2\ket{\alpha^{\prime 2}_{d-1}\omega_d'\alpha_d^{\prime 2}} & \ldots \\
        \iddots & \vdots & \ddots & \iddots & \vdots & \ddots \\
    \end{pNiceArray}
    \right]
    \begin{pmatrix}
        \vdots\\
        C^{\omega_d}_{1, \alpha_{d-1}\alpha_{d}}\\
        \vdots\\
        \vdots\\
        C^{\omega_d}_{2, \alpha_{d-1}\alpha_{d}}\\
        \vdots
    \end{pmatrix} \\
    =
    \begin{pmatrix}
        \vdots \\
        \bra{\alpha_{d-1}^1\omega_d\alpha_d^1}\left(\ket{a_1}-\ket{B_1}\right) \\
        \vdots \\
        \vdots \\
        \bra{\alpha_{d-1}^2\omega_d\alpha_d^2}\left(\ket{a_2}-\ket{B_2}\right) \\
        \vdots \\
    \end{pmatrix}\mathrm{.}
    \label{eq:big_LS}
\end{multline}
A single block matrix in Eq.\,\eqref{eq:big_LS} contains all possible combinations of $\bra{\alpha_{d-1}^i\omega_d\alpha_d^i}\hat{D}_i\hat{D}_j\ket{\alpha^{\prime j}_{d-1}\omega_d'\alpha_d^{\prime j}}$. Fortunately, we do not need to compute the whole (block) matrix. Instead, we treat all tensors of the resulting tensor network individually to avoid computationally expensive contractions. We solve the whole linear system by solving it block-wise which is illustrated in Eq.\,\eqref{eq:ax=b} (cf. Fig.\,\ref{fig:nonlinear_Ax}(b)).
\begin{equation}
    \begin{pmatrix}
        \ddots & \vdots & \iddots \\
        \ldots & \bra{\alpha_{d-1}^i\omega_d\alpha_d^i}\hat{D}_i\hat{D}_j\ket{\alpha^{j}_{d-1}\omega_d\alpha_d^{j}} & \ldots \\
        \iddots & \vdots & \ddots \\
    \end{pmatrix}
    \begin{pmatrix}
        \vdots\\
        C^{\omega_d}_{i, \alpha_{d-1}\alpha_{d}}\\
        \vdots\\
    \end{pmatrix} \\
    =
    \begin{gathered}
    \includegraphics[]{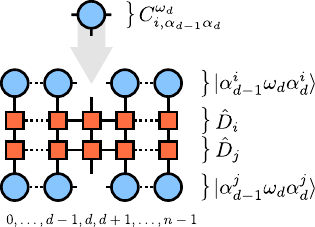}
    \end{gathered}
    \label{eq:ax=b}
\end{equation}
We employ the conjugate gradient method to obtain the solution of Eq.\,\eqref{eq:big_LS}, which corresponds directly to the updated tensor values of $\ket{v_1}$ and $\ket{v_2}$ in mixed canonical form (cf. Eq.\,\eqref{eq:mixed_canonical}). After updating the MPSs, we shift the canonical center to the neighbouring tensor to optimize its values. This procedure is repeated until we arrive at the first tensor again, which corresponds to a full optimization sweep. If the norm of the (unnormalized) MPS does not change, we interpret it as converged to its optimum and we save the velocity for the time step $s+1$. If we did not reach the final simulation time, we repeat the whole optimization loop to find the next time step. As an initial guess, we always take the velocity of the previous time step, as it should be close to next one.
\section{Initial conditions}\label{sec:initial}
In this paper, we have studied the decaying jet (DJ) flow \cite{gourianov_quantum-inspired_2022} and the decaying turbulence (DT) flow \cite{san_high-order_2012}. Both flows are defined on a square domain with periodic boundary conditions. For simplicity, the length of the square domain has length 1. This section provides the detailed definition of the initial conditions of DJ and DT flow, as shown in Fig.\,\ref{fig:problems} (a) and (c).

The DJ flow has initial conditions
\begin{equation}
    \mathbf{u}_\text{init}(x_1, x_2) = \mathbf{J}(x_2) + \mathbf{D}(x_1, x_2)
\end{equation}
with the jet profile
\begin{equation}
    \mathbf{J}(x_2) = \hat{\mathbf{e}}_1\frac{u_0}{2}\left[\tanh{\left(\frac{x_2-x_\text{min}}{h}\right)} - \tanh{\left(\frac{x_2-x_\text{max}}{h}\right)}-1\right]
\end{equation}
and disturbance
\begin{equation}
    \mathbf{D}(x_1, x_2) = \delta(\hat{\mathbf{e}}_1d_1(x_1, x_2)+\hat{\mathbf{e}}_2d_2(x_1, x_2))\mathrm{.}
\end{equation}
The subfunctions of $\mathbf{D}(x_1, x_2)$ are defined as
\begin{align}
\begin{split}
    d_1(x_1, x_2) =& \frac{2}{h^2}\left[(x_2-x_\text{max})e^{-(x_2-x_\text{max})^2/h^2}+(x_2-x_\text{min})e{-(x_2-x_\text{min})^2/h^2}\right]\\
    &\times\left[\sin(8\pi x_1)+\sin(24\pi x_1)+\sin(6\pi x_1)\right]\\
\end{split}
\end{align}
\begin{align}
    d_2(x_1, x_2) =& \pi\left[e^{-(x_2-x_\text{max})^2/h^2}+e^{-(x_2-x_\text{min})^2/h^2}\right]
    \times\left[8\cos(8\pi x_1)+24\cos(24\pi x_1)+6\cos(6\pi x_1)\right]\mathrm{.}
\end{align}
Here, $\delta=u_0/(40A)$, $A=\underset{x_1, x_2}{\max}\left(\sqrt{d_1(x_1, x_2)^2+d_2(x_1, x_2)^2}\right)$ and we have used the following values: $x_\text{min}=0.4$, $x_\text{max}=0.6$, $h=1/200$, and $u_0=1$. Further details are given in \cite{gourianov_exploiting_2022, gourianov_quantum-inspired_2022}.

The DT flow is initialized according to an initial spectral energy distribution
\begin{equation}
    E(\kappa) = \frac{a_s}{2}\frac{1}{\kappa_\text{p}}\left(\frac{\kappa}{\kappa_\text{p}}\right)^{2s+1}e^{-\left(s+1/2\right)(\kappa/\kappa_\text{p})^2}\mathrm{,}
\end{equation}
where $a_s = \frac{(2s+1)^{s+1}}{2^ss!}$ is a normalization coefficient. The parameter $s$ governs the shape of the energy distribution and $k_\text{p}$ defines the position of its peak. In our simulations, we set the values to $s=3$ and $k_\text{p}=7$. The vorticity in Fourier space is calculated as
\begin{equation}
    \hat{\omega}(\kappa_1, \kappa_2)=\sqrt{\frac{\kappa}{\pi}E(\kappa)}e^{i\zeta(k_1, k_2)}\mathrm{.}
\end{equation}
The function $\zeta(k_1, k_2)$ introduces a random global phase between $0$ and $2\pi$ for every point in wave space. We obtain the initial conditions by taking the Fourier transform of $\hat{\omega}(\kappa_1, \kappa_2)$ and inverting the relation $\omega = \nabla\times\mathbf{u}$. Further details and a rigorous study of DT flow is given in \cite{san_high-order_2012}.
\section{Direct Numerical Simulation Scheme}\label{sec:dns}
In this paper, we have compared quantum-inspired simulation (QIS) to direct numerical simulation (DNS). Here, we provide a brief description of the DNS scheme based on the Matlab implementation of \cite{vuorinen_dnslab_2016}. A similar DNS scheme has been used in \cite{gourianov_quantum-inspired_2022}.

As introduced in Sec.\,\ref{sec:ifd}, the Navier-Stokes equations consist of the continuity
\begin{equation}
    \nabla\cdot\bm{u} = 0
    \label{eq:continuityapp}
\end{equation}
and the momentum equation
\begin{equation}
    \frac{\partial}{\partial t}\bm{u} = \underbrace{-\left(\bm{u}\cdot\nabla\right)\bm{u}}_\text{convection} + \underbrace{\frac{1}{\text{Re}}\nabla^2 \bm{u}}_\text{diffusion} - \nabla p \text{.}
    \label{eq:momentumapp}
\end{equation}
These equations describe the motion of incompressible fluids and the DNS scheme yields a numerical solution. Therefore, we further divide the momentum equation into convection $\mathbf{C}$ and diffusion $\mathbf{D}$. Thus, we can simplify the momentum equation to
\begin{equation}
    \frac{\partial}{\partial t}\bm{u} = \mathbf{C} + \mathbf{D} -\nabla p\mathrm{.}
    \label{eq:momentumcd}
\end{equation}
By taking the divergence of Eq.\,\eqref{eq:momentumcd}, we can derive the pressure Poisson equation
\begin{equation}
    \Delta p = -\nabla\cdot\mathbf{C}\mathrm{.}
    \label{eq:pressurepoisson}
\end{equation}
This allows us to recover the pressure from the velocity field and is one of the basic principles of the so-called projection method. Hence, we can ignore the pressure term in Eq.\,\eqref{eq:momentumcd} and solve the Navier-Stokes equations first for an intermediate velocity $\mathbf{u^*}$. Using again the Euler method, we obtain the intermediate velocity for the subsequent time step by
\begin{equation}
    \mathbf{u^*}=\mathbf{u}^s-\Delta t(\mathbf{C}^s-\mathbf{D}^s)\mathrm{.}
\end{equation}
Here, $\mathbf{C}^s$ and $\mathbf{D}^s$ correspond to the convection and diffusion term evaluated at time step $s$ with $\mathbf{u}^s$. To ensure that the continuity equation is fulfilled, we need to project the intermediate velocity $\mathbf{u^*}$ into the divergence-free subspace:
\begin{equation}
    \mathbf{u}^{s+1} = \mathbf{u^*}-\nabla p\mathrm{.}
    \label{eq:helm}
\end{equation}
Eq.\,\eqref{eq:helm} is justified by the Helmholtz-theorem, splitting the smooth intermediate velocity in a sum of a divergence-free ($\mathbf{u}$) and a curl-free part ($\nabla p$). To perform this projection step, we require the pressure gradient. Therefore, we recover $p$ using the pressure Poisson equation \eqref{eq:pressurepoisson} to compute $\mathbf{u}^{s+1}$. 

Since the DJ and DT flow have periodic boundary conditions, we can perform the projection step in spectral space without computing pressure explicitly. This becomes clear by looking at the continuity equation \eqref{eq:continuityapp} in Fourier space
\begin{equation}
    i\boldsymbol{\kappa}\cdot\mathbf{\hat{u}}(\kappa_1, \kappa_2) = 0\mathrm{.}
\end{equation}
This is simply an orthogonality condition between wave vector $\boldsymbol{\kappa}$ and the Fourier coefficient $\mathbf{\hat{u}}(\kappa_1, \kappa_2)$. This orthogonality can be ensured using simple vector projection
\begin{equation}
    \mathbf{\hat{u}}^{s+1} = \mathbf{\hat{u}}^*-i\boldsymbol{\kappa}\frac{\mathbf{\hat{u}}^*\cdot\boldsymbol{\kappa}}{\kappa^2}\mathrm{.}
    \label{eq:projection}
\end{equation}

Summing up, the DNS scheme has the following steps:
\begin{enumerate}
    \item Calculate the intermediate velocity $\mathbf{u}^*$ by solving the momentum equation without pressure term\footnote{The momentum equation without pressure term is also known as Burger's equation.}.
    \item Compute the Fourier transform $\mathbf{\hat{u}}^*$.
    \item Project $\mathbf{\hat{u}}^*$ to divergence-free space by Eq.\,\eqref{eq:projection}.
    \item Compute the inverse Fourier transform to retrieve $\mathbf{\hat{u}}^{s+1}$.
\end{enumerate}
These steps are repeated until the desired time has been reached. This Euler scheme can be easily extended to the RK4 scheme.

To compute the derivatives in $\mathbf{C}$ and $\mathbf{D}$, we use central finite difference stencils with eighth order accuracy \cite{fornberg_generation_1988}. Moreover, we use the skew-symmetric form of the convection term to improve numerical stability (see Ref.\,\cite{vuorinen_dnslab_2016} for details).

\section{Comparison to Prior Work}\label{sec:discrepancy}
\begin{figure}[h]
    \includegraphics{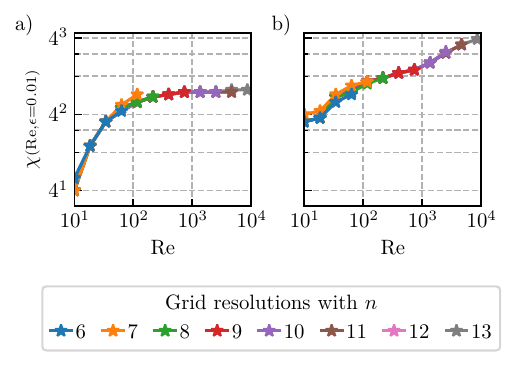}
    \caption{The impact of increasing Reynolds number $\mathrm{Re}$ on the required maximum bond dimension for a $\SI{99}{\percent}$ accurate MPS representation of the (a) x- and (b) y-component of the velocity.}
    \label{fig:chi_Re_scaling}
\end{figure}

In this section, we aim to provide a brief comparison with previous work in \cite{gourianov_quantum-inspired_2022} and explain the observed discrepancies. To this end, we examine DJ, as this flow problem has been analyzed in both studies. For the comparison, we divide our value for the Reynolds number by 200 ($\mathrm{Re}\rightarrow\mathrm{Re}/200$) to match the Reynolds number reported in the reference. Using DNS results, we examine the encoding efficiency similarily as described in Sec.\,\ref{sec:approx}. We then focus on the maximum bond dimension $\chi(\mathrm{Re}, \epsilon=0.01)$ to achieve a representation with an error of $\epsilon=0.01$. The results for various \(n\) are presented in Fig.\,\ref{fig:chi_Re_scaling}.
The same analysis for the similar problem is shown in Fig.\,1\,d) of \cite{gourianov_quantum-inspired_2022}. By comparing both figures, one can see that Gourianov et al. only considered $\chi(\mathrm{Re}, \epsilon=0.01)$ for the $x_1$-component of the flow. The maximum bond dimension of the $x_1$-component saturates earlier compared to the $x_2$-component in Fig.\,\ref{fig:chi_Re_scaling}\,b). This explains, why previously a $\chi_\mathrm{sat}(\epsilon=0.01)$ of 25 was reported. Our analysis differentiates between $\chi_\mathrm{sat}^x(\epsilon=0.01) = 25$ and $\chi_\mathrm{sat}^y(\epsilon=0.01) = 72$. However, within the considered range of Reynolds numbers between 10 and \(10^3\), no saturation of \(\chi(\epsilon=0.01)\) can be observed in our analysis. Instead, it appears that $\chi(\mathrm{Re}, \epsilon=0.01)$ follows a power law, similar to what is depicted in Fig.\,1\,(d) of \cite{gourianov_quantum-inspired_2022} for their 3D flow problem. Only due to our analysis for significantly higher Reynolds numbers, we know that $\chi(\mathrm{Re}, \epsilon=0.01)$ ultimately does saturate. This naturally raises the question of whether $\chi(\mathrm{Re}, \epsilon=0.01)$ for the 3D problem in \cite{gourianov_quantum-inspired_2022} might also saturate at higher Reynolds numbers or for different values of \(\epsilon\) or \(n\). Furthermore, Fig.\,\ref{fig:chi_Re_scaling} demonstrates that the bond dimension behaves independently of \(n\), as long as \(n\) is large enough for the simulation to run accurately on the \(2^n\times2^n\) grid.

\section{Derivation of the Scaling Law for the Maximum Bond Dimension $\chi$ with Error $\epsilon$}\label{sec:derivation_scaling}
To understand the origin of the power law of the TKE spectrum, we recap the definition of kinetic energy
\begin{equation}
    E(\kappa_1, \kappa_2) = \frac{1}{2} \left(\hat{u}_1(\kappa_1, \kappa_2)^2 + \hat{u}_2(\kappa_1, \kappa_2)^2\right)\mathrm{,}
\end{equation}
and the definition of enstrophy
\begin{equation}
    \mathcal{E}(\kappa_1, \kappa_2) = \frac{1}{2} \hat{\omega}(\kappa_1, \kappa_2)^2.
\end{equation}
Here, $\hat{\omega}(\kappa_1, \kappa_2)$ is the Fourier transform of the vorticity $\omega(x_1, x_2) = \nabla\times\bm{u}$. Both quantities are conserved in 2D flows\footnote{In 3D flows, enstrophy is not conserved \cite{pope_turbulent_2000}.} \cite{lesieur_turbulence_2008}. Due to these conservation laws, Batchelor argued that the energy spectrum of 2D decaying flows depends only on the wave number $\kappa$ and on the enstrophy dissipation rate $\beta_\text{ens}$ \cite{batchelor_computation_1969, lesieur_turbulence_2008}:
\begin{equation}
    E(\kappa) \propto \beta_\text{ens}^{a}\kappa^{b}\mathrm{.}
    \label{eq:dimensional}
\end{equation}
The parameters $a$ and $b$ can be determined by a dimensional analysis. We know the dimensions of every term in Eq.\,\eqref{eq:dimensional}:
\begin{equation}
    E(\kappa) = [L]^3[T]^{-2}\text{, } \beta_\text{ens}=[T]^{-3}\text{, }\kappa=[L]^{-1}\mathrm{.}
\end{equation}
Here, $[L]$ and $[T]$ are the dimensions of space and time, respectively. Thus, $a=2/3$ and $b=-3$.

As we know now that the TKE spectrum follows the power law $E(\kappa)\propto\kappa^{-3}$ in the inertial range, we can derive a similar power law for the Fourier coefficients of the velocity fluctuations $|\hat{u}'(\kappa_1, \kappa_2)|$. The TKE spectrum is calculated as
\begin{equation}
    E(\kappa) = \frac{1}{2} \int_0^{2\pi}|\hat{u}'(\kappa_1, \kappa_2)|^2 \kappa d\phi\mathrm{.}
    \label{eq:2djacobi}
\end{equation}
For simplicity, we consider the TKE spectrum only for a single spatial dimension. To calculate the total energy distribution, one would simply add up all energy spectra from each spatial dimension.
If we assume that $\hat{u}'(\kappa_1, \kappa_2) = C\kappa^{-\gamma}$, we calculate
\begin{align}
    E(\kappa) &= \frac{1}{2} \int_0^{2\pi} C^2 \kappa^{-2\gamma+1} d\phi\\
              &= C^2\pi\kappa^{-2\gamma+1}\mathrm{.}
\end{align}
Here, $C$ is some constant.
Thus, we know that $\gamma=2$ such that $E(\kappa)\propto\kappa^{-3}$.
Ref. \cite{jobst_efficient_2023} has shown that an image with algebraically decaying Fourier coefficients 
\begin{equation}
    |\hat{F}(\kappa_1, \kappa_2)|\leq C\frac{1}{(|\kappa_1|+1)^\alpha}\frac{1}{(|\kappa_2|+1)^\beta}
    \label{eq:desired}
\end{equation}
can be approximated with an MPS with maximum bond dimension $\chi=\mathcal{O}((1/\epsilon)^{1/(\text{min}(\alpha, \beta)-1/2)})$, where $\epsilon$ is the approximation error as defined in Eq.\,\eqref{eq:epsilon} and $\alpha,\beta>1$. Although we do not encode images but velocity fields as MPSs, we can directly use that result. As shown above, the Fourier coefficients of the velocity fluctuations decay as
\begin{equation}
    |\hat{u}'(\kappa_1, \kappa_2)| \leq C\kappa^{-2} = \frac{C}{\kappa_1^2+\kappa_2^2}\mathrm{.}
    \label{eq:v_power}
\end{equation}
As discussed in Sec.\,\ref{sec:derivation}, we assume that the Fourier coefficients of the pure velocity $\hat{u}(\kappa_1, \kappa_2)$ behave similarly for large $\kappa$, since the mean velocity should cover large-scale features, and thus, it should mainly have nonnegligible Fourier coefficients at low $\kappa$. Unfortunately, Eq.\,\eqref{eq:v_power} and Eq.\,\eqref{eq:desired} do not have the same form. Therefore, we extend Eq.\,\eqref{eq:v_power} by using the inequality of arithmetic and geometric means $\frac{x+y}{2}\geq\sqrt{xy}$, where $x, y$ are non-negative numbers:
\begin{equation}
    |\hat{u}(\kappa_1, \kappa_2)| \leq \frac{C}{2}\frac{1}{|\kappa_1|}\frac{1}{|\kappa_2|}\mathrm{.}
\end{equation}
Consequently, we have the expression of Eq.\,\eqref{eq:desired} with $\alpha=1$ and $\beta=1$\footnote{The denominators in Eq.\,\eqref{eq:desired} additionally contain "+1" to avoid a division by zero.}. However, the derivation of Ref. \cite{jobst_efficient_2023} assumes that $\alpha,\beta>1$. Actually, $\alpha$ and $\beta$ are very close to 1 but $\alpha,\beta>1$ is true, as the power law $E(\kappa)\propto\kappa^{-3}$ is an asymptotic result. In practice, the TKE spectrum shows a slightly steeper decay (cf. p.318 of \cite{lesieur_turbulence_2008}) with $E(\kappa)\propto\kappa^{-(3+\delta)}$ and $\delta>0$. Hence, $\alpha=\beta=1+\delta/4$ and we can use the result $\chi=\mathcal{O}((1/\epsilon)^{4/(2+\delta)})\approx\mathcal{O}((1/\epsilon)^{2})$. This bound is very conservative and in practice, we observe a less steep decrease with $\epsilon$ in Fig.\,\ref{fig:chi_sat}. Consequently, we relax the scaling to be of polynomial nature $\chi=\mathcal{O}(\text{poly}(1/\epsilon))$. For extremely large $\kappa$ and finite Reynolds number, the TKE spectrum is possibly exponentially decaying \cite{lesieur_turbulence_2008}, which is an even stronger argument to find a good MPS approximation. Ref. \cite{jobst_efficient_2023} also considers the case of exponentially decaying Fourier coefficients leading to $\chi=\mathcal{O}(\log(1/\epsilon))$. Since the inertial range takes up most of the spectrum and our data shows polynomial scaling of $\chi$ with $1/\epsilon$, we neglected the dissipation range for our analysis.

For 3D turbulence, the TKE spectrum scales as $E(\kappa)\propto\kappa^{-5/3}$. For the 3D case, the Jacobi determinant looks different than in Eq.\,\eqref{eq:2djacobi}. We have
\begin{equation}
    E(\kappa) = \frac{1}{2} \int_0^{\pi}\int_0^{2\pi}|\hat{u}'(\kappa_1, \kappa_2, \kappa_3)|^2 \sin\theta\kappa^2 d\phi d\theta\mathrm{.}
    \label{eq:2djacobi}
\end{equation}
If we assume again that $\hat{u}'(\kappa_1, \kappa_2, \kappa_3) = C\kappa^{-\gamma}$, we get
\begin{align}
    E(\kappa) &= \frac{1}{2} \int_0^{\pi}\int_0^{2\pi}C^2\kappa^{-2\gamma+2} \sin\theta d\phi d\theta\\
              &= 2C^2\pi\kappa^{-2\gamma+2}\mathrm{.}
\end{align}
Consequently, $\gamma=11/6$ and by using the extended inequality of arithmetic and geometric means $\frac{x+y+z}{3}\geq(xyz)^{1/3}$, we have
\begin{equation}
    |\hat{u}(\kappa_1, \kappa_2, \kappa_3)| \leq C\frac{1}{|\kappa_1|^{11/18}}\frac{1}{|\kappa_2|^{11/18}}\frac{1}{|\kappa_3|^{11/18}}\mathrm{.}
\end{equation}
Since $11/18<1$, we cannot use the same bound as before. However, the TKE spectrum also shows an exponential decay in the dissipation range for $\kappa\gg1$. Future investigations need to clarify, whether this exponential decay is sufficient for an efficient MPS approximation. This analysis should be confirmed by empirical data and maybe, one finds a tighter bound for $\chi$, where the Fourier coefficients scale as $\propto\kappa^{-11/6}$.
\end{document}